\documentclass[a4paper,10pt]{article}

\usepackage{amssymb}
\usepackage{amsmath}
\usepackage{subfigure}
\usepackage{slashed}
\usepackage{amsmath}

\allowdisplaybreaks
\usepackage{soul}
\usepackage{graphicx}
\usepackage{amsthm}
\usepackage{latexsym}
\usepackage[dvips]{epsfig}

\usepackage{wasysym}
\usepackage{mathrsfs}
\usepackage{eufrak}
\usepackage{bm}
\usepackage{authblk}
\usepackage{slashed}
\usepackage{yhmath} 
\usepackage{stmaryrd}

\theoremstyle{plain}
\newtheorem{proposition}{Proposition}
\newtheorem{lemma}{Lemma}
\newtheorem{theorem}{Theorem}
\newtheorem{assumption}{Assumption}

\newtheorem{remark}{Remark}

\def\bma{{\bm a}}
\def\bmb{{\bm b}}
\def\bmc{{\bm c}}
\def\bmd{{\bm d}}
\def\bme{{\bm e}}

\def\bmg{{\bm g}}
\def\bmh{{\bm h}}

\def\bml{{\bm l}}
\def\bmn{{\bm n}}
\def\bmm{{\bm m}}

\def\bmA{{\bm A}}
\def\bmB{{\bm B}}
\def\bmC{{\bm C}}
\def\bmD{{\bm D}}
\def\bmE{{\bm E}}

\def\bmeta{{\bm \eta}}

\def\bmsigma{{\bm \sigma}}

\def\bmpartial{{\bm \partial}}

\newcounter{mnotecount}%[section]

\newcommand{\mnotex}[1]%{}
{\protect{\stepcounter{mnotecount}}$^{\mbox{\footnotesize $\bullet$\themnotecount}}$ 
\marginpar{%\color{red}%
\raggedright\tiny\em
$\!\!\!\!\!\!\,\bullet$\themnotecount: #1} }

\usepackage[utf8]{inputenc}
\usepackage[english]{babel}
 
\usepackage{multicol}
\usepackage{color}
 
\usepackage{comment}

\usepackage[dvipsnames]{xcolor}
 
\begin{document}

\title{\textbf{ Zero rest-mass fields and the Newman-Penrose constants
    on flat space }}

\author[1,2]{E. Gasper\'in\footnote{E-mail address:{\tt
      edgar.gasperin-garcia@u-bourgogne.fr}}} \author[,3]{J. A. Valiente
  Kroon \footnote{E-mail address:{\tt j.a.valiente-kroon@qmul.ac.uk}}}

\affil[1]{Institut  de  Math\'ematiques  de  Bourgogne  (IMB),  UMR  5584,  CNRS, Universit\'e  de  Bourgogne  Franche-Comt\'e,  F-21000  Dijon,  France.}
  \affil[2] {CENTRA, Departamento de F\'isica, Instituto Superior
  T\'ecnico IST, Universidade de Lisboa UL, Avenida Rovisco Pais 1,
  1049 Lisboa, Portugal} \affil[3]{School of Mathematical Sciences,
  Queen Mary, University of London, Mile End Road, London E1 4NS,
  United Kingdom.}

\maketitle

\begin{abstract}
Zero rest-mass fields of spin 1 (the electromagnetic field) and spin 2
propagating on flat space and their corresponding Newman-Penrose (NP)
constants are studied near spatial infinity.  The aim of this analysis
is to clarify the correspondence between data for these fields on a
spacelike hypersurface and the value of their corresponding NP
constants at future and past null infinity. To do so, Friedrich's framework of
the cylinder at spatial infinity is employed to show that, expanding
the initial data in terms spherical harmonics and powers of the
geodesic spatial distance $\rho$ to spatial infinity, the NP constants
correspond to the data for the second highest
possible spherical harmonic at fixed order in $\rho$.
In addition, it is shown that for generic initial
  data within the class considered in this article, there is no
  natural correspondence between the NP constants
  at future and past null infinity ---for both the Maxwell and spin-2
  field. However, if the initial data is
  time-symmetric then the NP constants at
  future and past null infinity have the same information.
\end{abstract}



\section{Introduction}

The concept of asymptotic simplicity is central for the understanding
of isolated systems in general relativity. In this regard, Penrose's
proposal \cite{Pen63} is an attempt to characterise the fall-off
behaviour of the gravitational field in a geometric manner ---see also
\cite{Fri04}.  The essential mathematical idea behind for the Penrose
proposal is that of a \emph{conformal transformation}: given a
spacetime $(\tilde{\mathcal{M}},\tilde{\bmg})$ satisfying the Einstein
field equations (\emph{the physical spacetime}) one considers a
4-dimensional Lorentzian manifold $\mathcal{M}$ equipped with a metric
$\bmg$ such that $\bmg$ and $\tilde{\bmg}$ are conformal to each
other, in other words
\[
\bmg=\Xi^2 \tilde{\bmg},
\]
where $\Xi$ is the so-called \emph{conformal factor}. The pair
$(\mathcal{M},\bmg)$ can be called the \emph{unphysical spacetime}.
The set of points where $\Xi=0$ but
$\textbf{\mbox{d}}\Xi \neq 0$ is called the \emph{null infinity} and is
denoted by $\mathscr{I}$.  If $\tilde{\bmg}$ satisfies the vacuum
Einstein field equations (with vanishing Cosmological constant) near
$\mathscr{I}$, then the conformal boundary defines a smooth null
hypersurface of $\mathcal{M}$ ---see \cite{Fri04, Ste91}.  One can
identify two disjoint pieces of $\mathscr{I}$: $\mathscr{I}^{-}$ and
$\mathscr{I}^{+}$ correspond to the past and future end points of
null geodesics. If every null geodesic acquires two distinct endpoints
at $\mathscr{I}$, the spacetime $(\tilde{\mathcal{M}},\tilde{\bmg})$ is
said to be \emph{asymptotically simple} ---see \cite{Fri04, Ste91,
  CFEBook} for precise definitions.  The Minkowski spacetime,
$(\mathbb{R}^4, \tilde{\bmeta})$ is the prototypical example of an
asymptotically simple spacetime. In the standard conformal
representation of the Minkowski spacetime, the unphysical spacetime
 can be identified with the Einstein cylinder $(\mathcal{M}_{E},\bmg_{E})$
where $\mathcal{M}_{\mathcal{E}} \approx \mathbb{R} \times \mathbb{S}^3$ and
\[
\bmg_{E}= \mathbf{d}T\otimes \mathbf{d}T -\mathbf{d}\psi \otimes
\mathbf{d}\psi-\sin^{2}\psi\bmsigma , \qquad \Xi=\cos(T)+ \cos (\psi),
\]
 where $-\pi < T < \pi$, $0< \psi < \pi$ and $\bmsigma$ is the
 standard metric on $\mathbb{S}^2$. In this conformal representation
 $\mathscr{I}^{\pm}$ correspond to the sets of points on the Einstein
 cylinder, $\mathcal{M}_{\mathcal{E}}\equiv
 \mathbb{R}\times\mathbb{S}^3$, for which $0<\psi<\pi$ and $T=
 \pm(\pi-\psi)$. One can directly verify that
 $\Xi|_{\mathscr{I}^{\pm}}=0$ while
 $\textbf{\mbox{d}}\Xi|_{\mathscr{I}^{\pm}} \neq 0$ ---see
 \cite{Ste91}. Consequently, a distinguished region in the conformal
 structure of the Minkowski spacetime is \emph{spatial infinity}
 $i^{0}$ for which both $\Xi|_{i^{0}}$ and
 $\textbf{\mbox{d}}\Xi|_{i^{0}}$ vanish.  In this conformal
 representation, spatial infinity corresponds to a point in the
 Einstein cylinder with coordinates $\psi=\pi$ and $T=0$.

\medskip

A natural problem to be considered is the existence 
 of spacetimes whose conformal structure  
 resembles that of the Minkowski spacetime.  In this
 setting, the conformal Einstein field equations introduced originally
 in \cite{Fri81a} provide a convenient framework for
 discussing global existence of \emph{asymptotically simple} solutions
 to the Einstein field equations. An important application of these
 equations is the proof of the semi-global non-linear
 stability of the Minkowski spacetime given in \cite{Fri81a}. In the latter
 work, the evolution of perturbed initial data close to
 exact Minkowski data is analysed. Nevertheless, the 
initial data is not prescribed on a Cauchy hypersurface
 $\mathcal{\tilde{S}}$ but in an
 hyperboloid $\mathcal{\tilde{H}}$ whose  conformal extension
 in $\mathcal{M}$ intersects $\mathscr{I}$.
Therefore, an open problem in the framework of the conformal Einstein field
equations is the analysis of the evolution of initial data prescribed
on a Cauchy hypersurface $\mathcal{S}$ intersecting $i^{0}$ ---see
\cite{ChrKla93} for the proof of the global non-linear stability of the
Minkowski spacetime employing different methods.
  One of the main difficulties in
establishing a global result for the stability of the Minkowski
spacetime using conformal methods lies  on the fact that the
initial data for the conformal Einstein field equations is not 
smooth at $i^{0}$. This is not unexpected since, as 
observed by Penrose ---see \cite{Pen63, Pen65a}, the
conformal structure of spacetimes with non-vanishing mass
becomes singular at spatial infinity.  
 A milestone in the resolution of this problem is the
construction,  originally introduced in \cite{Fri98a},
 of a new representation of spatial
infinity known as the \emph{cylinder at spatial infinity}.
  In this representation, spatial infinity is not represented as a
  point but as set whose topology is that of a cylinder. This 
representation is well adapted to exploit the properties of
  curves with special conformal properties: \emph{conformal geodesics}.
In addition, it allows to formulate a regular finite initial value
problem for the conformal Einstein field equations ---other approaches
for analysing the gravitational field near spatial infinity using
different representations of spatial infinity have been also proposed
in literature ---see \cite{Sch81,BeiSch82,Bei84,Sch87}.  

\medskip
The framework of the cylinder at spatial infinity and its connection with the
conformal Einstein field equations have been
exploited in an analysis of the \emph{gravitational Newman-Penrose (NP)
constants} in \cite{FriKan00}.  The NP constants, originally introduced
in \cite{NewPen68}, are defined in terms of integrals over cuts
$\mathcal{C} \approx \mathbb{S}^2$ of $\mathscr{I}$.  The integrands
in the expressions defining the NP constants are, however, written in
a particular gauge adapted to $\mathscr{I}$ (the so-called \emph{NP-gauge})
while the natural gauge used in the framework of the cylinder at
spatial infinity (the so-called \emph{F-gauge} in \cite{FriKan00}), is
adapted to a congruence of conformal geodesics and hinged at a Cauchy
hypersurface $\mathcal{S}$.  This fact, which in first instance looks
as an obstacle to analyse the NP constants, turns out to be
advantageous since, once the relation between the NP-gauge and the
F-gauge is clarified, one can relate the initial data prescribed on
$\mathcal{S}$ with the gravitational NP constants at $\mathscr{I}$.

\medskip

In \cite{HawPerStr16}, the authors exploited the notion of
these conserved quantities at $\mathscr{I}$ to make inroads into the
problem of the information paradox ---see \cite{Haw76,Haw74,Haw75}.
In the latter work, the concept of \emph{soft hair} is motivated by
means of an analysis of the conservation laws and symmetries of
abelian gauge theories in Minkowski space. These conservation laws
correspond essentially to the electromagnetic version of the
gravitational NP constants.  With this motivation, in the present
article zero rest-mass fields propagating on flat space and their
corresponding NP constants are studied.  Two physically relevant
fields are analysed: the spin-1 and spin-2 zero rest-mass fields. The
spin-1 field provides a description of the electromagnetic field while the
spin-2 field on the Minkowski spacetime describes linearised gravity.

 In this article it is shown how the framework of the cylinder at
 spatial infinity can be exploited to relate the corresponding NP
 constants with the initial data on a Cauchy hypersurface intersecting
 $i^{0}$ ---see Propositions
 \ref{PropositionNPconstantsGeneralStructure} and
 \ref{PropositionNPconstantsGeneralStructureTimeDual} for the spin-1
 case and Proposition
 \ref{PropositionSpin2NPconstantsGeneralStructure} and
 \ref{PropositionSpin2NPconstantsGeneralStructureTimeDual} for the
 spin-2 case. Additionally, it is shown that, for
   generic initial data within the class considered, the NP constants
   at $\mathscr{I}^{+}$ and $\mathscr{I}^{-}$ do not coincide.
   However, the correspondence between the NP constants at
   $\mathscr{I}^{+}$ and $\mathscr{I}^{-}$ is fulfilled for
   time-symmetric initial data ---see Theorems
   \ref{TheoremEMconstants} and \ref{TheoremSpin2constants}.

\subsection{Outline of the paper}

Section \ref{CylinderAtspatialInfinity} contains a general discussion
of the cylinder at spatial infinity and the F-gauge in the Minkowski
spacetime. In Section \ref{TheMaxwellEquationsInTheFGauge}, the
Maxwell equations are written in the F-gauge and the initial data for
the electromagnetic field on a spacelike hypersurface is discussed.
In Section \ref{LinearisedGravityEquations}, the equations governing the
massless spin-2 field are expressed in the F-gauge and 
 the corresponding initial data is discussed.
In Section \ref{sec:NPGauge}, Bondi coordinates and a NP-frame for a
conformal extension of the Minkowski spacetime is derived.
Additionally, the relation between this frame and the one introduced
in Section \ref{CylinderAtspatialInfinity} is determined explicitly.
In Section \ref{TheElectromagneticNPConstants} the electromagnetic NP
constants are introduced and written in the F-gauge. This
construction is exploited to identify the electromagnetic NP constants
with part of the initial data introduced in Section
\ref{TheMaxwellEquationsInTheFGauge}. In Section 
\ref{LinearisedGravityNPConstants}
a similar analysis is carried out for the spin-2 field; the
corresponding NP constants are found and written in terms of the initial data.
  Section \ref{Conclusions} provides with some concluding remarks.  In
  addition, a general discussion of the connection on $\mathbb{S}^2$
  is given in Appendix \ref{ConnectionTwoSphere} and a discussion of
  the $\eth$ and $\bar{\eth}$ operators of Newman and Penrose is
  provided in Appendix \ref{AppendixEth}.

\subsection{Notations and Conventions}

 The signature convention for (Lorentzian) spacetime metrics will be $
 (+,-,-,-)$.  In the rest of this article $\{_a ,_b , _c , . . .\}$
 denote abstract tensor indices and $\{_\bma ,_\bmb , _\bmc , . . .\}$
 will be used as spacetime frame indices taking the values ${ 0,
   . . . , 3 }$.  In this way, given a basis $\{\bme_{\bma}\}$ a
 generic tensor is denoted by $T_{ab}$ while its components in the
 given basis are denoted by $T_{\bma \bmb}\equiv
 T_{ab}\bme_{\bma}{}^{a}\bme_{\bmb}{}^{b}$.  Part of the analysis will
 require the use of spinors.  In this respect, the notation and
 conventions of Penrose \& Rindler \cite{PenRin84} will be followed.
 In particular, capital Latin indices $\{ _A , _B , _C , . . .\}$ will
 denote abstract spinor indices while boldface capital Latin indices
 $\{ _\bmA , _\bmB , _\bmC , . . .\}$ will denote frame spinorial
 indices with respect to a specified spin dyad ${ \{\delta_\bmA{}^{A}
   \} }.$
The  conventions for the curvature tensors are fixed by the relation
\[
(\nabla_a \nabla_b -\nabla_b \nabla_a) v^c = R^c{}_{dab} v^d.
\]

\section{The cylinder at spatial infinity and the F-Gauge}
\label{CylinderAtspatialInfinity}

In this section a conformal representation of the Minkowski spacetime
that is adapted to a congruence of conformal geodesics is
discussed. This conformal representation, introduced originally in
\cite{Fri98a}, is particularly suited for analysing the behaviour of
fields near spatial infinity. In broad terms, in this representation
spatial infinity $i^{0}$, which corresponds to a point in the standard
compactification of the Minkowski spacetime, is blown up to a
two-sphere $\mathbb{S}^2$. In the subsequent discussion 
this representation will be referred 
 as the \emph{cylinder at spatial infinity}. The
discussion of the cylinder at spatial infinity as presented in
\cite{Fri98a} is given in the language of fibre bundles. In particular,
the construction of the so-called \emph{extended bundle space} is
required ---see \cite{Fri98a,AceVal11}. Nevertheless, a discussion
which does not make use of this construction is presented in the
following.

\subsection{The cylinder at spatial infinity}
\label{TheCylinderAtSpatialInfinity}
Consider the Minkowski metric $\tilde{\bmeta}$ in cartesian
coordinates $\tilde{x}^{\alpha }= (\tilde{t},\tilde{x}^{i})$,
 \[\tilde{\bmeta}=\eta_{\mu\nu}\mathbf{d}\tilde{x}^{\mu}\otimes\mathbf{d}\tilde{x}^{\nu},\]
 where $\eta_{\mu\nu}=\text{diag}(1,-1,-1,-1)$. Introducing polar
 coordinates defined by $\tilde{\rho}=
 \delta_{ij}\tilde{x}^{i}\tilde{x}^{j}$ where
 $\delta_{ij}=\text{diag(1,1,1)}$, and an arbitrary choice of
 coordinates on $\mathbb{S}^2$, the metric $\tilde{\bmeta}$ can be
 written as
\begin{equation}\label{MinkowskiMetricPhysicalPolar}
\tilde{\bmeta}=\mathbf{d}\tilde{t}\otimes\mathbf{d}\tilde{t}
-\mathbf{d}\tilde{\rho}\otimes \mathbf{d}\tilde{\rho}-\tilde{\rho}^2
\mathbf{\bm\sigma},
\end{equation}
with $\tilde{t}\in(-\infty)$, $\tilde{\rho}\in [0,\infty)$ and
  $\bm\sigma$ denotes the standard metric on $\mathbb{S}^2$.  A common
  procedure to obtain a conformal representation of the Minkowski
  spacetime close to $i^{0}$ is to introduce \emph{inversion coordinates}
  $x^{\alpha}=(t,x^{i})$ defined by ---see
  \cite{Ste91},
 \[ 
x^{\mu}=-{\tilde{x}^{\mu}}/{\tilde{X}^2}, \qquad \tilde{X}^2 \equiv
 \tilde{\eta}_{\mu\nu}\tilde{x}^{\mu}\tilde{x}^{\nu}.
\]
 The inverse transformation is given by
 \[\tilde{x}^{\mu}=-x^{\mu}/X^2, \qquad
 X^2=\eta_{\mu\nu}x^{\mu}x^{\nu}.\] 
Using these coordinates one
 readily identifies the following conformal representation 
of the Minkowski spacetime
\begin{equation}
\label{InverseMinkowskiMetricDef}
\bmg_{I}=\Xi^2 \hspace{0.5mm}\tilde{\bmeta},
\end{equation}
where $\bmg_{I}=\eta_{\mu\nu}\mathbf{d}x^{\mu}\otimes
\mathbf{d}x^{\nu}$ and $\Xi =X^2$. Notice, additionally that,
$X^2=1/\tilde{X}^2$. Introducing an \emph{unphysical polar coordinate}
defined as $\rho=\delta_{ij}x^{i}x^{j}$, one observes that the
rescaled metric $\bmg_{I}$ and conformal factor $\Xi$ read
\begin{equation}
\label{InverseMinkowskiUnphysicaltrhocoords}
\bmg_{I}=\mathbf{d}t\otimes\mathbf{d}t -\mathbf{d}\rho\otimes
\mathbf{d}\rho-\rho^2 \mathbf{\bm\sigma}, \qquad \Xi=t^2-\rho^2,
\end{equation}
with $t\in(-\infty,\infty)$ and $\rho\in (0,\infty)$.
In this conformal representation, spatial infinity $i^{0}$ corresponds
to  a point located at the origin.
For future reference, observe that $\tilde{t}$ and $\tilde{\rho}$
are related to $t$ and $\rho$ via
\begin{equation}
\label{UnphysicalToPhysicaltrho}
\tilde{t}=-\frac{t}{t^2-\rho^2}, \qquad  \tilde{\rho}= \frac{\rho}{t^2-\rho^2}.
\end{equation}
 Then, one introduces a time coordinate $\tau$ defined via
 $t=\rho\tau$. In the coordinate system determined by $\tau$ and
 $\rho$ the metric $\bmg_{I}$ is written as
\begin{equation*}
\bmg_{I}=\rho^2 \mathbf{d}\tau\otimes \mathbf{d}\tau
-(1-\tau^2)\mathbf{d}\rho \otimes \mathbf{d}\rho + \rho\tau
\mathbf{d}\rho\otimes \mathbf{d}\tau + \rho\tau \mathbf{d}\tau \otimes
\mathbf{d}\rho - \rho^2 \bmsigma.
\end{equation*}
The required conformal representation is obtained by considering the
rescaled metric
\begin{equation}
\label{MetricCylinderToIversedMinkowski}
\bmg \equiv \frac{1}{\rho^2} \bmg_{I}.
\end{equation}

  Explicitly one has that 
\begin{align}\label{ExplicitFormMetricCylinder}
\bmg=\mathbf{d}\tau \otimes \mathbf{d}\tau
-\frac{(1-\tau^2)}{\rho^2}\mathbf{d}\rho \otimes \mathbf{d}\rho
+ \frac{\tau}{\rho}(\mathbf{d}\tau \otimes \mathbf{d}\rho + \mathbf{d}\rho \otimes \mathbf{d}\tau) -\bm\sigma.
\end{align}

Observe that  spatial infinity $i^{0}$, which is at infinity 
 respect to the metric $\bmg$, corresponds to a set
which has the topology of $\mathbb{R}\times\mathbb{S}^2$ ---see
\cite{Fri98a, AceVal11}.  In what follows we continue using the
coordinates $(\tau,\rho)$ and call them the  \emph{F-coordinates}.  Following the conformal rescalings previously
introduced one considers the conformal extension $(\mathcal{M},\bmg)$ where
\begin{equation}\label{MetricAndConformalFactor}
\bmg=\Theta^2 \tilde{\bm\eta}, \qquad \Theta=\rho(1-\tau^2),
\end{equation}
and 
\[
\mathcal{M} \equiv \{ p \in \mathbb{R}^4 \; \rvert \; -1 \leq \tau \leq 1 , \; \; \rho(p)\geq 0\}.
\]
In this representation future and past null infinity are
located at
\[
 \mathscr{I}^{+} \equiv \{ p \in \mathcal{M} \; \rvert\; \tau(p) =1 \}, 
\qquad \mathscr{I}^{-} \equiv \{ p \in \mathcal{M} \; \rvert \; 
 \tau(p) =-1\},
\]
and the physical Minkowski spacetime can be
identified with the region
\[
\tilde{\mathcal{M}} \equiv \{ p \in \mathcal{M} \; \rvert \; -1<\tau(p)<1 , \; \;\rho(p)>0 \},
\]
In addition, the following sets will be distinguished:
\begin{equation*}
 I \equiv \{ p \in \mathcal{M} \; \rvert   \;\;  |\tau(p)|<1, \; \rho(p)=0\}, 
\qquad I^{0} \equiv \{ p \in \mathcal{M}\; \rvert \;
  \tau(p)=0, \; \rho(p)=0\},
\end{equation*} 
\begin{equation*}
 I^{+} \equiv \{ p\in \mathcal{M} \; \rvert \; \tau(p)=1, \; \rho(p)=0
  \}, \qquad I^{-} \equiv \{p \in \mathcal{M}\; \rvert \; \tau(p)=-1, \; \rho(p)=0\}.
\end{equation*}
Notice that spatial infinity $i^0$, which originally was a point in the
$\bmg_{I}-$representation, can be identified with the set $I$ in the
$\bmg-$representation. In addition, one can intuitively think of
the \emph{critical sets} $I^{+}$ and $I^{-}$ as the region where
spatial infinity ``touches'' $\mathscr{I}^{+}$ and $\mathscr{I}^{-}$
respectively. Similarly, $I^{0}$ represents the intersection of
$i^{0}$ and the initial hypersurface $\mathcal{S} \equiv
\{\tau=0\}$. See \cite{Fri98a, FriKan00} and \cite{AceVal11} for
further discussion of the framework of the cylinder at spatial
infinity implemented for stationary spacetimes.

\subsection{The F-gauge}
\label{SectionF-gauge}

In this section a brief discussion of the so-called F-gauge is
provided ---see \cite{FriKan00,AceVal11} for a discussion of the
F-gauge in the language of fibre bundles. Following the philosophy of
the previous section, the discussion presented here will not make use
of the extended bundle space ---see \cite{FriKan00,AceVal11} for definitions.
  One of
the motivations for the introduction of this gauge is that it exploits
the properties of conformal geodesics. More precisely, in this
framework, one introduces a null frame whose timelike leg corresponds
to the tangent of a conformal geodesic starting from a fiduciary
spacelike hypersurface $\mathcal{S}=\{\tau=0\}$.  The notion of
conformal geodesics, however, will not be discussed here 
---see \cite{Fri95,Fri03c,Tod02,CFEBook}
for definitions and further discussion.

\medskip
 
To start the discussion, consider the conformal extension
$(\mathcal{M},\bmg)$ of the Minkowski spacetime and the
F-coordinate system introduced in Section \ref{TheCylinderAtSpatialInfinity}.
 Observe
that the induced metric on the surface $ \mathcal{Q}\equiv\{
\tau=\tau_{\star},\rho=\rho_{\star}, \}$, with $\tau_{\star},
\rho_{\star}$ fixed, is the standard metric on $\mathbb{S}^2$.
Consequently, one can
introduce a complex null frame $\{ \bm\partial_{+},\bm\partial_{-}\}$
on $\mathcal{Q}$ as described in Appendix
\ref{ConnectionTwoSphere}. To propagate this frame off
$\mathcal{Q}$ one requires that
\[
[\bm\partial_{\tau},\bm\partial_{\pm}]=0, \qquad
[\bm\partial_{\rho},\bm\partial_{\pm}]=0.
\]
Taking into account the above construction one writes, in spinorial
notation, the following spacetime frame
\begin{subequations}\label{Fframe}
\begin{align}
&\bme_{\bm0\bm0'}=\frac{\sqrt{2}}{2}\big((1-\tau)\bm\partial_{\tau} +
\rho\bm\partial_{\rho}\big), \qquad 
\bme_{\bm1\bm1'}=\frac{\sqrt{2}}{2}\big((1+\tau)\bm\partial_{\tau} -
\rho\bm\partial_{\rho}\big), \label{Fframe0011}\\
&\bme_{\bm0\bm1'}=\frac{\sqrt{2}}{2}\bm\partial_{+}, \qquad \qquad
 \qquad \qquad \quad
\bme_{\bm1\bm0'}=\frac{\sqrt{2}}{2}\bm\partial_{-}. \label{Fframe0110}
\end{align}
\end{subequations}
The corresponding dual coframe is given by
\begin{align*}
&\bm\omega^{\bm0\bm0'}=\frac{\sqrt{2}}{2}\Big( \mathbf{d}\tau
-\frac{1}{\rho}\big(1-\tau\big)\mathbf{d}\rho\Big),\qquad
\bm\omega^{\bm1\bm1'}=\frac{\sqrt{2}}{2}\Big(\mathbf{d}\tau +
\frac{1}{\rho}\big(1+\tau\big)\mathbf{d}\rho\Big),\\
&\bm\omega^{\bm0\bm1'}=\sqrt{2}\bm\omega^{+}, \qquad \qquad \qquad \qquad 
\qquad \quad \bm\omega^{\bm1\bm0'}=\sqrt{2}\bm\omega^{-}.
\end{align*}
One can directly verify that
\[
\bmg=\epsilon_{\bmA
  \bmB}\epsilon_{\bmA'\bmB'}\bm\omega^{\bmA\bmA'}\bm\omega^{\bmB\bmB'}.
\]
The above construction and frame will be referred in the following
discussion as the \emph{F-gauge}. A
direct computation using the Cartan structure equations
 shows that the only non-zero reduced connection
coefficients are given by
\begin{align*}
&\Gamma_{\bm0\bm0'}{}^{\bm1}{}_{\bm1}=\Gamma_{\bm1\bm1'}{}^{\bm1}{}_{\bm1}
=\frac{\sqrt{2}}{4},
  \qquad
  \Gamma_{\bm0\bm0'}{}^{\bm0}{}_{\bm0}=\Gamma_{\bm1\bm1'}{}^{\bm0}{}_{\bm0}
=-\frac{\sqrt{2}}{4},
  \\ &\Gamma_{\bm1\bm0'}{}^{\bm1}{}_{\bm1}=-\Gamma_{\bm1\bm0'}{}^{\bm0}{}_{\bm0}
=\frac{\sqrt{2}}{4}\omega,
  \qquad
  \Gamma_{\bm0\bm1'}{}^{\bm0}{}_{\bm0}=-\Gamma_{\bm0\bm1'}{}^{\bm1}{}_{\bm1}
=\frac{\sqrt{2}}{4}\overline{\omega}.
\end{align*}

\section{The electromagnetic field in the F-gauge}
\label{TheMaxwellEquationsInTheFGauge}

In this section the Maxwell equations on $(\mathcal{M},\bmg)$ are
discussed.  After rewriting the equations in terms of the $\eth$ and
$\bar\eth$ operators, a general solution is obtained by expanding the
fields in spin-weighted spherical harmonics. The resulting equations
for the coefficients of the expansion, satisfy ordinary differential
equations which can be explicitly solved in terms of special
functions. The analysis given here is similar to the one for the Maxwell
field on a Schwarzschild background in \cite{Val07b} and the
gravitational field in \cite{Fri98a}. Notice that, in contrast with
the analysis presented in this section, in the latter references the
equations and relevant structures are lifted to the extended bundle
space. Additionally, the initial data considered in this analysis is
generic and in particular is not assumed to be time-symmetric.

\subsection{The spinorial Maxwell equations}
\label{TheSpin-1Equation}

The Maxwell equations in the 2-spinor formalism take the form of the
spin-1 equation
\begin{equation}
\label{Spin-1Equation}
\nabla_{A'}{}^{A}\phi_{AB}=0.
\end{equation}
Let $\epsilon_{\bmA}{}^{A}$ with $\epsilon_{\bm0}{}^{A}=o^A$
 and $\epsilon_{\bm1}{}^{A}=\iota^A$
 denote a spin dyad adapted to the F-gauge so that
$e_{\bmA\bmA'}{}^{AA'}=\epsilon_{\bmA}{}^{A}\epsilon_{\bmA'}{}^{A}$,
corresponds to the null frame introduced in Section
\ref{SectionF-gauge}.  A direct computation shows that equation
\eqref{Spin-1Equation} implies  a set of equations for the
components of $\phi_{AB}$ respect to $\epsilon_{\bmA}{}^{A}$:
$\phi_{0}\equiv\phi_{AB}o^{A}o^{B}$,
$\phi_{1}\equiv\phi_{AB}o^{A}\iota^{B}$  and
$\phi_{2}\equiv\phi_{AB}\iota^{A}\iota^{B}$, which can
be split into a system of evolution equations
\begin{subequations}
\begin{align}
& (1+\tau)\bm\partial_{\tau}\phi_{0}-\rho\bm\partial_{\rho}\phi_{0}
  -\bm\partial_{+}\phi_{1}=-\phi_{0}, \label{EvolutionMaxwellEq0}\\ &
  \bm\partial_{\tau}\phi_{1}-\frac{1}{2}\big(\bm\partial_{+}\phi_{2}
  +\bm\partial_{-}\phi_{0}\big)=\frac{1}{2}\big(\overline{\varpi}\phi_{2}+
  \varpi\phi_{0}\big),\label{EvolutionMaxwellEq1}\\ &
  (1-\tau)\bm\partial_{\tau}\phi_{2} +\rho\bm\partial_{\rho}\phi_{2}
  -\bm\partial_{-}\phi_{1}=\phi_{2},\label{EvolutionMaxwellEq2}
\end{align}
 and a constraint equation 
\begin{equation}
 \tau\bm\partial_{\tau}\phi_{1}-\rho\bm\partial_{\rho}\phi_{1}
  +\frac{1}{2}\big(\bm\partial_{-}\phi_{0}
  -\bm\partial_{+}\phi_{2}\big)=\frac{1}{2}\big(\overline{\varpi}\phi_{2}
  - \varpi\phi_{0}\big).\label{ConstraintMaxwell}
\end{equation}
\end{subequations}

%%%%%%%%%%%%%%%%%%%%%%%

One can systematically solve the above equations decomposing the
fields $\phi_{0}, \phi_{1},\phi_{2}$ in spin-weighted spherical
harmonics. To do so, one has to rewrite these equations in terms
of the $\eth$ and $\bar{\eth}$ operators of Newman and Penrose.
Using \eqref{DirectionalDerivativesToEthandEthbar} of Appendix 
\ref{AppendixEth} and the fact that
$\phi_{0},\phi_{1}$ and $\phi_{2}$ have spin weights 1, 0 and -1,
respectively, one finds that equations
\eqref{EvolutionMaxwellEq0}-\eqref{ConstraintMaxwell} can be rewritten
as the following evolution equations
\begin{subequations}
\begin{align}
& (1+\tau)\bm\partial_{\tau}\phi_{0}-\rho\bm\partial_{\rho}\phi_{0} +
  \eth\phi_{1}=-\phi_{0}, \label{EthEvolutionMaxwellEq0}\\ &
  \bm\partial_{\tau}\phi_{1} + \frac{1}{2}\big(\eth\phi_{2}
  +\bar{\eth}\phi_{0}\big)= 0, \label{EthEvolutionMaxwellEq1}\\ &
  (1-\tau)\bm\partial_{\tau}\phi_{2} +\rho\bm\partial_{\rho}\phi_{2}
  +\bar{\eth}\phi_{1}=\phi_{2},\label{EthEvolutionMaxwellEq2}
\end{align}
and the constraint equation
\begin{equation}
 \tau\bm\partial_{\tau}\phi_{1}-\rho\bm\partial_{\rho}\phi_{1}
 +\frac{1}{2}\big(\eth\phi_{2}
 -\bar{\eth}\phi_{0}\big)=0.\label{EthConstraintMaxwell}
\end{equation}
\end{subequations}

\subsection{The transport equations
for the electromagnetic field
 on the cylinder at spatial infinity}
\label{TransportEqns}

In order  to analyse the behaviour of solutions of the
Maxwell equations in a neighbourhood of the cylinder at spatial
infinity it will be assumed
that $\phi_{0},\phi_{1}$ and $\phi_{2}$ are smooth functions
of $\tau$ and $\rho$. Moreover, taking into account equation
\eqref{GeneralExpansionSphericalHarmonics} of Appendix \ref{AppendixEth}
the following Ansatz is made:

\begin{assumption}
{\em The
components of the Maxwell field admit a Taylor-like expansion around
$\rho=0$ of the form
\begin{equation}
 \phi_{n}= \sum_{p=|1-n|}^{\infty}\sum_{\ell=|1-n|}^{p}\sum_{m=-\ell}^{\ell}
  \frac{1}{p!}a_{n,p;\ell,m}(\tau)Y_{1-n;\ell, m}
  \rho^{p}, \label{ExpansionPhin}
\end{equation} 
where $a_{n,p;\ell m}:\mathbb{R}\rightarrow \mathbb{C}$ and with
$n=0,1,2$. }
\end{assumption}

 \begin{remark}
\label{Remark:JustifyingExpansions}
 { \em
 For the purposes pursued in this article the expansion
 \eqref{ExpansionPhin} is understood as an Ansatz for the
 solution. Nevertheless, the structure of the expansion
 \eqref{ExpansionPhin} can be motivated from analysing the
 electromagnetic constraint equations. The formal nature of the above
 Ansatz can be controlled making use of the theory developed in
 \cite{Fri03b} ---see also \cite{Val07b,Val09a}. This, in turn, allows
 to find conditions on the freely specifiable initial
 data for the Maxwell equations ensuring the existence of
 expansions of the from given by \eqref{ExpansionPhin}. Obtaining such
 conditions, however, goes beyond the scope of the present article and
 will be discussed elsewhere.
 }
 \end{remark}

 To simplify the notation of the subsequent analysis let 
\begin{equation}\label{NotationPhinCoef}
\phi_{n}^{(p)} \equiv\frac{\partial^{p}\phi_{n}}{\partial
  \rho^{p}}\Bigg|_{\rho=0},
\end{equation}
 with $n=0,1,2$. Formally differentiating
equations \eqref{EthEvolutionMaxwellEq0}-\eqref{EthConstraintMaxwell}
respect to $\rho$ and evaluating at the cylinder $I$ one obtains
\begin{subequations}
\begin{align}
& (1+\tau)\dot{\phi}_{0}^{p}-(p-1)\phi_{0}^{p} +
  \eth\phi^{p}_{1}=0, \label{pthDerivativeEthEvolutionMaxwellEq0}\\ &
  \dot{\phi}^{(p)}_{1} + \frac{1}{2}\big(\eth\phi^{(p)}_{2}
  +\bar{\eth}\phi^{(p)}_{0}\big)=
  0, \label{pthDerivativeEthEvolutionMaxwellEq1}\\ &
  (1-\tau)\dot{\phi}^{(p)}_{2} + (p-1)\phi^{(p)}_{2}
  +\bar{\eth}\phi^{(p)}_{1}=0,\label{pthDerivativeEthEvolutionMaxwellEq2}\\ &
  \tau\dot{\phi}^{(p)}_{1}-p\phi^{(p)}_{1}
  +\frac{1}{2}\big(\eth\phi^{(p)}_{2}
  -\bar{\eth}\phi^{(p)}_{0}\big)=0,
\label{pthDerivativeEthConstraintMaxwell}
\end{align}
\end{subequations}
where the dot denotes a derivative respect to $\tau$.  Using equations
\eqref{RiseSpin}-\eqref{LowerSpin} of Appendix \ref{AppendixEth} and expansions encoded 
in equation \eqref{ExpansionPhin} one obtains the following
equations for $a_{n,p;\ell m}$
\begin{align}
&(1+\tau)\dot{a}_{0,p;\ell m} + \sqrt{\ell(\ell + 1)} a_{1,p;\ell
    m}-(p-1)a_{0,p;\ell m}=0,\label{EqCoefficient0Evolution}
  \\ &\dot{a}_{1,p;\ell m} + \frac{1}{2} \sqrt{\ell(\ell +
    1)}(a_{2,p;\ell m}-a_{0,p;\ell m})=0,
\label{EqCoefficient1Evolution}\\
&(1-\tau)\dot{a}_{2,p;\ell m}-\sqrt{\ell(\ell+1)}a_{1,p;\ell m} +
(p-1)a_{2,p;\ell,m}=0, \label{EqCoefficient2Evolution}\\ & \tau
\dot{a}_{1,p;\ell m}- \frac{1}{2}\sqrt{\ell(\ell + 1)}(a_{2,p;\ell m}
+ a_{0,p;\ell m})-p a_{1,p;\ell m}=0,\label{EqCoefficient1Constraint}
\end{align}
for $p \geq 1$, $1 \leq\ell\leq p$, $-\ell \leq m \leq \ell$.  Notice that equations
\eqref{EqCoefficient0Evolution}-\eqref{EqCoefficient1Constraint}
correspond, essentially, to the homogeneous part of the equations reported
in \cite{Val07b}. Furthermore, $a_{1,p;\ell,m}$ can be solved from
\eqref{EqCoefficient1Evolution} and \eqref{EqCoefficient1Constraint}
in terms of $a_{0,p;\ell m}$ and $a_{2,p;\ell,m}$ to obtain
\begin{equation}
\label{a1-coefficient}
a_{1,p;\ell m}=\frac{\sqrt{\ell(\ell + 1)}}{2p}
\big((1-\tau)a_{2,p;\ell.m} + (1+\tau)a_{0,p;\ell,m}\big).
\end{equation}
Substituting $a_{1,p;\ell,m}$ as given in \eqref{a1-coefficient} into
equations \eqref{EqCoefficient0Evolution} and
\eqref{EqCoefficient2Evolution} one obtains
\begin{subequations}
\begin{align}
& (1+\tau)\dot{a}_{0,p;\ell,m} + \Big(\frac{1}{2p}\ell(\ell+1)(1+\tau)
  -(p -1)\Big)a_{0,p;\ell,m}+\frac{1}{2p}\ell(\ell +
  1)(1-\tau)a_{2,p;\ell m}=0,
\label{FundamentalSystem1}\\
& (1-\tau)\dot{a}_{2,p;\ell m} -\frac{1}{2p}\ell(\ell +
1)(1+\tau)a_{0,p;\ell m} -\Big(\frac{1}{2p}\ell(\ell+1)(1-\tau) -( p -
1)\Big)a_{2,p;\ell m} =0.
\label{FundamentalSystem2}
\end{align}
\end{subequations}
At this point one can proceed in analogous way as in \cite{Val07b} to
obtain a fundamental matrix for the system
\eqref{FundamentalSystem1}-\eqref{FundamentalSystem2}:
a direct computation shows that 
one can decouple the last system of first order equations and obtain
the following second order equations
\begin{subequations}
\begin{align}
&(1-\tau^2)\ddot{a}_{0,p;\ell,m} + 
2(1-(1-p)\tau)\dot{a}_{0,p;\ell,m}+ 
(p+\ell)(\ell-p+1)a_{0,p;\ell,m}=0,
\label{SecondOrderEqCoef0}\\
&(1-\tau^2)\ddot{a}_{2,p;\ell,m}
 - 2(1 + (1-p)\tau)\dot{a}_{2,p;\ell,m} 
+ (p+\ell)(\ell-p+1)a_{2,p;\ell,m}=0.
\label{SecondOrderEqCoef2}
\end{align}
\end{subequations}
Dropping temporarily the subindices $p,\ell,m$ observe that, if
$a_{2}(\tau)$ solves \eqref{SecondOrderEqCoef2} then
$a_{2}^{s}(\tau)\equiv a_{2}(-\tau)$ solves equation
\eqref{SecondOrderEqCoef0}.  Equations
\eqref{SecondOrderEqCoef0}-\eqref{SecondOrderEqCoef2} are particular
examples of so-called \emph{Jacobi ordinary differential
  equations}. Following the discussion of \cite{Val07b} one obtains
the following:

  \begin{proposition}\label{PropSolutionsEMcase}
For $p \geq 1$, $ \ell < p$, $-\ell \leq m \leq \ell$ the solutions to
the Jacobi equations
\eqref{SecondOrderEqCoef0}-\eqref{SecondOrderEqCoef2} are polynomial
in $\tau$. Moreover,
\begin{equation*}
  \begin{pmatrix}
    a_{0,p;l,m}(\tau)  \\
    a_{2,p;l,m}(\tau)
  \end{pmatrix}
  =
  X_{p,\ell}(\tau) \left(X_{p,\ell}^{-1}(0)
    \begin{pmatrix}
    a_{0,p;l,m}(0) \\
    a_{2,p;l,m}(0) 
    \end{pmatrix}
    \right),
\end{equation*}
 where the fundamental matrix is given by
\begin{equation*}
    X_{p,\ell}(\tau) =
  \begin{pmatrix}
    Q^1_{p,\ell}(\tau) & (-1)^{\ell+1}Q^3_{p,\ell}(\tau) \\ (-1)^{\ell+1}Q_3(-\tau) &
    Q^1_{p,\ell}(-\tau) \\
  \end{pmatrix},
  \end{equation*}  
\begin{align*}
  &\text{with} & Q^1_{p,\ell}(\tau)=\Big(\frac{1-\tau}{2}\Big)^{p+1}P_{\ell-1}^{p+1,-p+1}(\tau) \qquad  & Q^3_{p,\ell}(\tau)=
  \Big(\frac{1+\tau}{2}\Big)^{p-1}P_{\ell+1}^{-p-1,p-1}(\tau).
\end{align*}
  \end{proposition}
  For future identification of the NP constants in terms of initial data
  it is   convenient to introduce some
  notation at this point. The matrix $X^{-1}_{p,l}(0)$
  is a symmetric matrix whose components will be represented as
  \begin{equation}
    X^{-1}_{p,l}(0) =
    \begin{pmatrix}
      X^{-1}_A & X^{-1}_B \\
      X^{-1}_B & X^{-1}_A
      \end{pmatrix}.
  \end{equation}
  The explicit values of $X^{-1}_A$, $X^{-1}_B$ can be determined by
inverting the fundamental matrix and direct evaluation the the Jacobi
polynomials at $\tau=0$.  The only relevant feature for the subsequent
discussion is that $X^{-1}_A \neq X^{-1}_B$.  Expanding the matrix
expression given in the above proposition one obtains:

  \begin{lemma}\label{MaxwellSolutionsExplicit}
    The solutions of Proposition \ref{PropSolutionsEMcase} can be
    written as:
\begin{align*}
  a_{0,p;l,m}(\tau) & = C_{p,\ell,m}Q^1_{p,\ell}(\tau) +
  (-1)^{1+\ell}D_{p,\ell,m}Q^3_{p,\ell}(\tau), \\ a_{2,p;l,m}(\tau)& =
  D_{p,\ell,m}Q^1_{p,\ell}(-\tau) +
  (-1)^{1+\ell}C_{p,\ell,m}Q^3_{p,\ell}(-\tau),
\end{align*}
\begin{align*}
 & where& C_{p,\ell,m} \equiv & X^{-1}_Aa_{0,p;l,m}(0) + X^{-1}_B
  a_{2,p;l,m}(0), & D_{p,\ell,m} \equiv & X^{-1}_Ba_{0,p;l,m}(0) +
  X^{-1}_A a_{2,p;l,m}(0).
\end{align*}
  \end{lemma}

Proposition \ref{PropSolutionsEMcase} encodes the solution for $p>l$. The
  $p=l$ is a special case for which one has the following:
\begin{proposition}\label{MaxwellLogarithmicTerms}
For  $p\geq 1$, $\ell=p$, $-p \leq m \leq p$ one has:
\begin{subequations}
\begin{align}
a_{0,p;p,m}(\tau)= \left(\frac{1-\tau}{2}\right)^{p+1}
\left(\frac{1+\tau}{2}\right)^{p-1} \left( E_{p,m}+
 E^{\ast}_{p,m}\int_{0}^{\tau}\frac{\mbox{ds}}{(1+s)^{p}(1-s)^{p+2}} \right),
\label{MaxwellSolution0Log}\\
a_{2,p;p,m}(\tau)= \left(\frac{1+\tau}{2}\right)^{p+1}
\left(\frac{1-\tau}{2}\right)^{p-1} \left( I_{p,m}+
 I^{\ast}_{p,m}\int_{0}^{\tau}\frac{\mbox{ds}}{(1-s)^{p}(1+s)^{p+2}} \right)
\label{MaxwellSolution2Log},
\end{align}
\end{subequations}
where  $E_{p,m}$, $E^{\ast}_{p,m}$  and $I_{p,m}$, $I^{\ast}_{p,m}$  are integration 
constants. 
\end{proposition}

\begin{remark}
{\em For non-vanishing $E^{\ast}_{p,m}$
and $I^{\ast}_{p,m}$ the solutions $a_{0,p;p,m}(\tau)$ and
$a_{2,p;p,m}(\tau)$ with $p \geq 1$, $-p\leq m \leq p$,
 contain terms which diverge logarithmically near $\tau=\pm 1$.
}
\end{remark}

\subsection{Initial data for the Maxwell equations}
\label{InitialDataMaxwellsEqns}

Evaluating the constraint equation \eqref{EthConstraintMaxwell}
at $\tau=0$ gives the following equation 
\begin{equation}
\rho\bm\partial_{\rho}\phi_{1}
 -\frac{1}{2}\big(\eth\phi_{2}
 -\bar{\eth}\phi_{0}\big)=0.\label{InitialDataConstraint}
\end{equation}
Consistent with the expressions encoded in equation \eqref{ExpansionPhin}
one considers on the initial hypersurface
 $\mathcal{S}$ fields $\phi_{n}|_{\mathcal{S}}$, with $n=0,1,2$,
which can be expanded as
\begin{equation}
\phi_{n}|_{\mathcal{S}}= \sum_{p=|1-n|}^{\infty}\sum_{\ell=|1-n|}^{p}\sum_{m=-\ell}^{\ell}
  \frac{1}{p!}a_{n,p;\ell,m}(0)Y_{1-n;\ell, m}
  \rho^{p}, \label{InitialExpansionPhin}
\end{equation}

Observe that once $a_{0,p;\ell,m}(0)$ and $a_{2,p;\ell,m}(0)$ are
given, $a_{1,p;\ell,m}(0)$ is already determined by virtue of equation
\eqref{a1-coefficient} as
\begin{equation*}
a_{1,p;\ell m}(0)=\frac{\sqrt{\ell(\ell + 1)}}{2p}
\big(a_{2,p;\ell,m}(0) + a_{0,p;\ell,m}(0)\big)
\end{equation*}
In addition, observe that equations
\eqref{FundamentalSystem1}-\eqref{FundamentalSystem2} are first order
while equations \eqref{SecondOrderEqCoef0}-\eqref{SecondOrderEqCoef2}
are second order. Consequently, the initial data
$\dot{a}_{0,p,\ell,m}(0)$ and $\dot{a}_{2,p,\ell,m}(0)$ are
determined, by virtue of equations
\eqref{FundamentalSystem1}-\eqref{FundamentalSystem2} restricted to
$\mathcal{S}$, by the initial data $a_{0,p,\ell,m}(0)$ and
$a_{2,p,\ell,m}(0)$.

\medskip

  \begin{remark}\label{remarkGeneralSolutionToTheConstraints}
    \em{
    { Although identifying $a_{0,p,\ell,m}(0)$ and $a_{2,p,\ell,m}(0)$
as the free specifiable initial data is sufficient for the purposes of
this article, a systematic and geometric way to parametrise the
initial data is to write the solution to the constraint equation
\eqref{EthConstraintMaxwell} in terms of Hertz potentials.  Exploiting
the results of \cite{AndBacJodi14} one has that the general solution
to the constraint equation \eqref{EthConstraintMaxwell} can be written
as
\[\phi_{AB}= (\mathcal{G}_{2}\varphi)_{AB},\] where
$\varphi_{AB}=\varphi_{(AB)}$ is a symmetric but otherwise arbitrary
spinor encoding the free specifiable data. The operator
$\mathcal{G}_2$ is defined as $(\mathcal{G}_{2}\varphi)_{AB} \equiv
D_{(A}{}^{Q}\varphi_{B)Q}$ where $D_{AB}$ denotes the spinorial
counterpart of the Levi-Civita connection associated with metric
$\bmh$ intrinsic to the hypersurface $\mathcal{S}$.  The details of
this construction are not necessary for the main discussion of this
article and will be presented elsewhere.}}
\end{remark}

\begin{assumption}\label{RegularityConditionMaxwell}
  \em{ Although general initial data allows for solutions with
    $E^\star_{p,m} \neq 0$ and $I^\star_{p,m} \neq 0$ for the
    calculation of the NP constants it will be assumed that}
  \[
E^\ast_{p,m}=I^\ast_{p,m} =0.
  \]
  In other words, it will be assumed that the fields $\phi_{n}$ do not
  contain the diverging terms of Proposition
  \ref{MaxwellLogarithmicTerms}. 
\end{assumption}

\begin{remark}\label{TimeSymmetricInitialDataMaxwellMagneticPartVanishes}
  \em{
  The spin-1 field $\phi_{AB}$ (Maxwell spinor) can be decomposed into its
  electric and magnetic parts, denoted as $\eta_{AB}$ and $\mu_{AB}$
  respectively, as follows:
  \[
    \phi_{AB} = \eta_{AB}+\mbox{i}\mu_{AB},
  \]
  with
  \[
\eta_{AB} = \frac{1}{2}(\phi_{AB} + \phi^{\dagger}_{AB}), \qquad
\mu_{AB}=-\frac{1}{2}\mbox{i} \;(\phi_{AB} - \phi^{\dagger}_{AB}).
  \]
  where $\phi^{\dagger}_{AB} \equiv   \tau_{A}{}^{A'}\tau_{B}{}^{B'}\bar{\phi}_{A'B'}$ with
  $\tau_{A}{}^{A'} \equiv o_A o^{A'} + \iota_A \bar{\iota}^{A'}$.
  Here $\tau^{AA'}$ corresponds to the spinorial counterpart
  of the vector $\tau^a = \sqrt{2}\bm\partial_\tau$  ---see
  \cite{CFEBook} for further discussion on the space spinor formalism.
  Initial data for which $\mu_{AB}|_{\mathcal{S}}=0$ will be called
  time-symmetric. A calculation then shows that time-symmetric data
  satisfy,
  \[
\phi_0 = \bar{\phi}_2, \qquad \phi_1 = -\bar{\phi}_1 \qquad \text{on}
\quad \mathcal{S}.
  \]
  The latter conditions imply that  $a_{0,p,\ell,m}(0)=a_{2,p,\ell,m}(0)$.
  }
\end{remark}
  
  \begin{remark}\label{TimeSymmetricInitialData}
    { \em  If the data is time-symmetric then
       $C_{p,\ell,m}=D_{p,\ell,m}$. Nevertheless, for generic initial
       data one has $C_{p,\ell,m} \neq D_{p,\ell,m}$.}
  \end{remark}

\begin{remark}
{\em The convergence of the expansions encoded in \eqref{ExpansionPhin}
 follows from the results of \cite{Val09a}.}
\end{remark}

\section{The massless spin-2 field equations in the F-gauge}
\label{LinearisedGravityEquations}

In Section \ref{TheMaxwellEquationsInTheFGauge} 
the Maxwell equations (in the F-gauge) were discussed, these
correspond in spinorial formalism to the spin-1 equations. In this
section, a similar analysis is performed but now
 for a spin-2 field propagating
on the Minkowski spacetime. As
discussed in \cite{Val03a} the spin-2 equations
 where the background geometry is that of the  Minkowski spacetime can
be used to describe the linearised
gravitational field. In \cite{Val03a} these equations were written
in terms the lifts of the relevant structures to the extended bundle
space. In this section, following the spirit of the present article, the
equations will be discussed without making use of these structures. 
In a similar way as in the electromagnetic case studied in Section
 \ref{TheMaxwellEquationsInTheFGauge},
after rewriting the equations in terms of the $\eth$ and $\bar\eth$
operators, a general solution is obtained by expanding the fields in
spin-weighted spherical harmonics. The resulting equations for the
coefficients of the expansion satisfy ordinary differential equations
which can be explicitly solved in terms of special functions.

\subsection{The spin-2 equation}
\label{TheSpin2Equation}
As discussed in \cite{Val03a},
the linearised gravitational field over the Minkowski spacetime can be
described with the so-called
 massless spin-2 field equation
\begin{equation}\label{Spin2Equation}
\nabla_{A'}{}^{A}\phi_{ABCD}=0.
\end{equation}
Following an approach  analogous to the one described in Section 
\ref{TheSpin-1Equation}
 for the
electromagnetic field, it can be shown that
 equation \eqref{Spin2Equation}
implies the following  evolution equations for the components of the spinor
$\phi_{ABCD}$ 
\begin{subequations}
\begin{align}
& (1+\tau)\bm\partial_{\tau}\phi_{0}-\rho\bm\partial_{\rho}\phi_{0}-
  \bm\partial_{+}\phi_{1} +
  \bar{\varpi}\phi_{1}=-2\phi_{0}, \label{GravEq0}\\ &
  \bm\partial_{\tau}\phi_{1}-\frac{1}{2}\bm\partial_{+}\phi_{2}-
  \frac{1}{2}\bm\partial_{-}\phi_{0}
  -\varpi\phi_{0}=-\phi_{1}, \label{GravEq1}\\ &
  \bm\partial_{\tau}\phi_{2}-\frac{1}{2}\bm\partial_{-}\phi_{1}-\frac{1}{2}
\bm\partial_{+}\phi_{3}-
\frac{1}{2}\varpi\phi_{1}-\frac{1}{2}\bar{\varpi}\phi_{3}=0,
 \label{GravEq2}
  \\ &
  \bm\partial_{\tau}\phi_{3}-\frac{1}{2}\bm\partial_{+}
\phi_{4}-\frac{1}{2}\bm\partial_{-}\phi_{2}-
\bar{\varpi}\phi_{4}=\phi_{3}, \label{GravEq3}\\ &
  (1-\tau)\bm\partial_{\tau}\phi_{4}+ \rho
  \bm\partial_{\rho}\phi_{4}-\bm\partial_{-}\phi_{3}+
  \varpi\phi_{3}=2\phi_{4}, \label{GravEq4}
\end{align}
\end{subequations}
 and the constraint equations
\begin{subequations}
\begin{align}
&
  \tau\bm\partial_{\tau}\phi_{1}-\rho\bm\partial_{\rho}
\phi_{1}-\frac{1}{2}\bm\partial_{+}\phi_{2}+
  \frac{1}{2}\bm\partial_{-}\phi_{0}+
  \varpi\phi_{0}=0, \label{GravEq5}
\\ &
  \tau\bm\partial_{\tau}\phi_{2}-\rho\bm
\partial_{\rho}\phi_{2}-\frac{1}{2}\bm\partial_{+}\phi_{3}   +
  \frac{1}{2}\bm\partial_{-}\phi_{1}-\frac{1}{2}
\bar{\varpi}\phi_{3}+
  \frac{1}{2}\varpi \phi_{1}=0, \label{GravEq6}
 \\ &
  \tau\bm\partial_{\tau}\phi_{3}-\rho\bm\partial_{\rho}\phi_{3}
-\frac{1}{2}\bm\partial_{+}\phi_{4}   +
  \frac{1}{2}\bm\partial_{-}\phi_{2}-\bar{\varpi}\phi_{4}=0,
 \label{GravEq7}
\end{align}
\end{subequations}
where the five components $\phi_{0},\phi_{1},\phi_{2},\phi_{3}$ 
and $\phi_{4}$, given by 
\begin{eqnarray*} 
& \phi_{0}\equiv\phi_{ABCD}o^{A}o^{B}o^{C}o^{D}, \qquad
  \phi_{1}\equiv\phi_{ABCD}o^{A}o^{B}o^{C}\iota^{D}, \\ &
  \phi_{2}\equiv\phi_{ABCD}o^{A}o^{B}\iota^{C}\iota^{D}, \qquad
  \phi_{3}\equiv\phi_{ABCD}o^{A}\iota^{B}\iota^{C}\iota^{D},\\ &
  \phi_{4}\equiv\phi_{ABCD}\iota^{A}\iota^{B}\iota^{C}\iota^{D}, \qquad
  \phantom{\phi_{0}=\phi_{ABCD}o^{A}o^{B}o^{C}o^{D}}
\end{eqnarray*}
have spin weight of $2,1,0,-1,-2$ respectively.
Taking into account this observation and equations
 \eqref{DirectionalDerivativesToEthandEthbar}  and \eqref{EthAndEthBarForUs}
 given in Appendix \ref{AppendixEth} one can rewrite \eqref{GravEq0}-\eqref{GravEq7} in terms
of the $\eth$ and $\bar{\eth}$ as done for the electromagnetic case.
A direct computation renders the following evolution equations
\begin{subequations}
\begin{align}
& (1+\tau)\bm\partial_{\tau}\phi_{0}-\rho\bm\partial_{\rho}\phi_{0} +
  \eth\phi_{1}=-2\phi_{0}, \label{EvolutionGravEq0}\\ &
  \bm\partial_{\tau}\phi_{1} + \frac{1}{2} \bar{\eth}\phi_{0}+
  \frac{1}{2}\eth\phi_{2}=-\phi_{1}, \label{EvolutionGravEq1}\\ &
  \bm\partial_{\tau}\phi_{2} +\frac{1}{2}\bar{\eth}\phi_{1}+\frac{1}{2}
\eth \phi_{3}=0,
 \label{EvolutionGravEq2}
  \\ &
  \bm\partial_{\tau}\phi_{3}+\frac{1}{2}\bar{\eth}\phi_{2}+\frac{1}{2}\eth\phi_{4}
 =\phi_{3}, \label{EvolutionGravEq3}\\ &
  (1-\tau)\bm\partial_{\tau}\phi_{4}+ \rho
  \bm\partial_{\rho}\phi_{4}+ \bar{\eth} \phi_{3} = 2\phi_{4},
 \label{EvolutionGravEq4}
\end{align}
\end{subequations}
and the constraint equations
\begin{subequations}
\begin{align}
&  \tau\bm\partial_{\tau}\phi_{1}-\rho\bm\partial_{\rho}\phi_{1} + \frac{1}{2}
 \eth\phi_{2} - \frac{1}{2}\bar{\eth}\phi_{0} =0, \label{ConstraintGravEq5}
\\ &
  \tau\bm\partial_{\tau}\phi_{2}-\rho\bm\partial_{\rho}\phi_{2}
 + \frac{1}{2}\eth \phi_{3} - \frac{1}{2}\bar{\eth}\phi_{1} =0,
 \label{ConstraintGravEq6}
 \\ &
  \tau\bm\partial_{\tau}\phi_{3}-\rho\bm\partial_{\rho}\phi_{3}
+ \frac{1}{2}\eth \phi_{4}- \frac{1}{2}\bar{\eth}\phi_{2}=0. 
\label{ConstraintGravEq7}
\end{align}
\end{subequations}

With the equations already written in this way, 
one can follow the discussion of \cite{Val03a}
for parametrising the solutions to equations 
\eqref{EvolutionGravEq0}-\eqref{ConstraintGravEq7}.

\subsection{The transport equations for the
massless spin-2 field on the cylinder at spatial infinity}
\label{GravTransport}

One proceeds in analogous way as in the electromagnetic case and assumes
that the fields $\phi_{n}$ with $n = 0,1,2,3,4,$ are smooth functions
of $\tau$ and $\rho$. Taking into account equation
\eqref{GeneralExpansionSphericalHarmonics} of Appendix \ref{AppendixEth},
 it is assumed that one can express the
components the of the linearised gravitational field in a Taylor-like
expansion around $\rho=0$. More precisely, the following Ansatz is made: 

\begin{assumption}
In what follows it will be assumed that the components of the spin-2 field have
the expansions
\begin{equation}
 \phi_{n}= \sum_{p=|2-n|}^{\infty}\sum_{\ell=|2-n|}^{p}\sum_{m=-\ell}^{\ell}
 \frac{1}{p!}a_{n,p;\ell,m}(\tau)Y_{2-n;\ell,m}
 \rho^{p} \label{ExpansionGravPhin}
\end{equation}
where $a_{n,p;\ell,m}:\mathbb{R}\rightarrow\mathbb{C}$ and $n=0,\ldots,4$. 
\end{assumption}

\begin{remark}
{\em As in the case of the Maxwell equations (cf. Remark
  \ref{Remark:JustifyingExpansions}), these above Ansatz for the spin-2
field can be expressed in terms of condition on the initial data. This
analysis falls beyond the scope of the present article and 
will be discussed elsewhere.}
\end{remark}

For the remaining part of this section, 
the p-th derivative respect  to $\rho$ of the fields
 $\phi_{n}$  with $n=0,1,2,3,4$ evaluated  at the cylinder $I$,
is denoted using the
 same notation as in equation \eqref{NotationPhinCoef}.
Then, by formally differentiating equations \eqref{EvolutionGravEq0}-
\eqref{ConstraintGravEq7} 
respect to $\rho$ and
evaluating at the cylinder $I$, one obtains the following equations
\begin{subequations}
\begin{align}
& (1+\tau)\bm\partial_{\tau}\phi_{0}^{(p)} +
  \eth\phi_{1}^{(p)} (p-2)\phi_{0}^{(p)}=0, \label{pthDerivativeGravEq0}\\ &
  \bm\partial_{\tau}\phi_{1}^{(p)} + \frac{1}{2} \bar{\eth}\phi_{0}^{(p)}+
  \frac{1}{2}\eth\phi_{2}^{(p)} + \phi_{1}^{(p)}=0, \label{pthDerivativeGravEq1}\\ &
  \bm\partial_{\tau}\phi_{2} +\frac{1}{2}\bar{\eth}\phi_{1}^{(p)}+\frac{1}{2}
\eth \phi_{3}^{(p)}=0,
 \label{pthDerivativeGravEq2}
  \\ &
  \bm\partial_{\tau}\phi_{3}+\frac{1}{2}\bar{\eth}\phi_{2}^{(p)}
+\frac{1}{2}\eth\phi_{4}^{(p)}-\phi_{3}^{(p)}=0, \label{pthDerivativeGravEq3}\\ &
  (1-\tau)\bm\partial_{\tau}\phi_{4}^{(p)} + \bar{\eth} \phi_{3}^{(p)}  
+ (p- 2)\phi_{4}^{(p)}=0, \label{pthDerivativeGravEq4} 
\end{align}
\end{subequations}
and
\begin{subequations}
\begin{align}
&  \tau\bm\partial_{\tau}\phi_{1} + \frac{1}{2} \eth\phi_{2}^{(p)} - \frac{1}{2}\bar{\eth}\phi_{0}^{(p)}-p\phi_{1}^{(p)} =0, \label{pthDerivativeGravEq5}
\\ &
  \tau\bm\partial_{\tau}\phi_{2}
 + \frac{1}{2}\eth \phi_{3}^{(p)} - 
\frac{1}{2}\bar{\eth}\phi_{1}^{(p)}-p\phi_{2}^{(p)} =0, \label{pthDerivativeGravEq6}
 \\ &
  \tau\bm\partial_{\tau}\phi_{3}
+ \frac{1}{2}\eth \phi_{4}^{(p)}- \frac{1}{2}\bar{\eth}\phi_{2}^{(p)} 
-p\phi_{3}^{(p)}=0. 
\label{pthDerivativeGravEq7}
\end{align}
\end{subequations}
The last set of equations along with the expansion 
\eqref{ExpansionGravPhin}, in turn, imply 
the following  equations for $a_{n,p;\ell,m}$ with
 $p \geq 2$ and $2 \leq \ell \leq p$:
\begin{subequations}
\begin{align}
& (1+\tau)\dot{a}_{0} + \lambda_{1} a_{1} - (p-2) a_{0} =0,
 \label{CoefGravEq0} \\ &
  \dot{a}_{1} - \frac{1}{2} \lambda_{1} a_{0} +
  \frac{1}{2} \lambda_{0} a_{2} + a_{1} =0,
 \label{CoefGravEq1}\\ &
  \dot{a}_{2} -\frac{1}{2}\lambda_{0} a_{1} +\frac{1}{2}
\lambda_{0}a_{3}=0,
 \label{CoefGravEq2}
  \\ &
  \dot{a}_{3} - \frac{1}{2} \lambda_{0}a_{2}
+\frac{1}{2}\lambda_{1}a_{4} -a_{3}
=0, \label{CoefGravEq3}\\ &
  (1-\tau)\dot{a}_{4} - \lambda_{1}a_{3} 
+ (p- 2)a_{4}=0, \label{CoefGravEq4} 
\end{align}
\end{subequations}
and
\begin{subequations}
\begin{align}
&  \tau \dot{a}_{1} + \frac{1}{2} \lambda_{0}a_{2} + \frac{1}{2} 
\lambda_{1}a_{0}-p a_{1} =0, \label{CoefGravEq5}
\\ &
  \tau \dot{a}_{2}
 + \frac{1}{2}\lambda_{0}a_{3} + 
\frac{1}{2}\lambda_{0}a_{1}-p a_{2} =0, \label{CoefGravEq6}
 \\ &
  \tau \dot{a}_{3}
+ \frac{1}{2}\lambda_{1}a_{4} + \frac{1}{2} \lambda_{0}a_{2} 
-p a_{3}=0,
\label{CoefGravEq7}
\end{align}
\end{subequations}
where $\lambda_{1} \equiv \sqrt{(\ell-1)(\ell+2)}$ and $\lambda_{0}
\equiv \sqrt{\ell(\ell+1)}$ and the labels $p;\ell,m$ have been
suppressed for conciseness.  From equations
\eqref{CoefGravEq1}-\eqref{CoefGravEq3} and
\eqref{CoefGravEq5}-\eqref{CoefGravEq7} one obtains an algebraic
system which can be written succinctly as
\begin{equation}\label{AlgebraicSysGrav}
\left(
\begin{matrix}
 p+\tau & -\frac{1}{2}(1-\tau)\lambda_{0} & 0 \\[6pt]
 -\frac{1}{2}(1+\tau)\lambda_{0}  & p & -\frac{1}{2}(1-\tau)\lambda_{0} \\[6pt]
0 & -\frac{1}{2}(1+\tau)\lambda_{0} & p-\tau 
\end{matrix}
\right) \left( 
\begin{matrix}
a_{1}\\[6pt] a_{2} \\[6pt] a_{3}
\end{matrix}
\right) = \frac{1}{2}\lambda_{1}\left(
\begin{matrix}
(1+\tau) a_{0} \\[6pt] 0\\[6pt] (1-\tau) a_{4}
\end{matrix} \right).
\end{equation}
Solving the above system and substituting $a_{0}$, $a_{1}$ and $a_{3}$
written in terms of $a_{0}$ and $a_{4}$ into equations
 \eqref{CoefGravEq0} and \eqref{CoefGravEq4} 
one obtains
\begin{subequations}
\begin{eqnarray}
(1+\tau)\dot{a}_{0} + (-(p-2) + f(\tau,p,\ell) )a_{0} + g(\tau,p,\ell)a_{4}=0,
\label{CoupledFirstOrderGrav0}  \\
(1-\tau)\dot{a}_{4} + (-(p-2) + f(-\tau,p,\ell) )a_{4} + g(-\tau,p,\ell)a_{0}=0, 
\label{CoupledFirstOrderGrav4}
\end{eqnarray}
\end{subequations}
where
\begin{align*}
 & f(\tau,p,\ell) \equiv \frac{(1+\tau)(\ell-1)(\ell+2)[4p^2-4p\tau+
      \ell(\ell+1)(\tau^2-1)]}{4p(2p^2-\ell(\ell+1)+
    (\ell-1)(\ell+2)\tau^2)}, \\ &
  g(\tau,p,\ell) \equiv
  \frac{(1-\tau)^3\ell(\ell+1)
(\ell-1)(\ell+2)}{4p(2p^2-\ell(\ell+1)+(\ell-1)(\ell+2)\tau^2)}. 
\end{align*}
Together, the last equations entail the following decoupled equations
\begin{subequations}
\begin{eqnarray}
& (1-\tau^2)\ddot{a}_{0}+(4+ 2(p-1)\tau)\dot{a}_{0} + (p+ \ell)(p-\ell
  + 1)a_{0}=0,
\label{SecondOrderGrav0} \\
& (1-\tau^2)\ddot{a}_{4}+(-4+ 2(p-1)\tau)\dot{a}_{4} + (p+
\ell)(p-\ell + 1)a_{4}=0.
\label{SecondOrderGrav4}
\end{eqnarray}
\end{subequations}
It can be verified that if $a_{0}(\tau)$ solves \eqref{SecondOrderGrav0} 
then $a_{0}^{s}(\tau) \equiv a_{0}(-\tau)$ solves equation
 \eqref{SecondOrderGrav4}. 
As in the electromagnetic case, these equations are Jacobi ordinary
differential equations. For the solutions to these equations one has
the following:

\begin{proposition}\label{PropSolutionsSpin2}
For $p \geq 2$, $ \ell < p$, $-\ell \leq m \leq \ell$ the solutions to
the Jacobi equations
\eqref{SecondOrderGrav0}-\eqref{SecondOrderGrav4} are polynomial
in $\tau$. Moreover,
\begin{equation*}
  \begin{pmatrix}
    a_{0,p;l,m}(\tau)  \\
    a_{4,p;l,m}(\tau)
  \end{pmatrix}
  =
  X_{p,\ell}(\tau) \left(X_{p,\ell}^{-1}(0)
    \begin{pmatrix}
    a_{0,p;l,m}(0) \\
    a_{4,p;l,m}(0) 
    \end{pmatrix}
    \right),
\end{equation*}
 where the fundamental matrix is given by
\begin{equation*}
    X_{p,\ell}(\tau) =
  \begin{pmatrix}
    Q^1_{p,\ell}(\tau) & (-1)^{\ell}Q^3_{p,\ell}(\tau) \\ (-1)^{\ell}Q_3(-\tau) &
    Q^1_{p,\ell}(-\tau) \\
  \end{pmatrix},
  \end{equation*}  
\begin{align*}
  &\text{with} & Q^1_{p,\ell}(\tau)=\Big(\frac{1-\tau}{2}\Big)^{p+2}P_{\ell-2}^{p+2,-p+2}(\tau), \qquad
  & Q^3_{p,\ell}(\tau)=  \Big(\frac{1+\tau}{2}\Big)^{p-2}P_{\ell+2}^{-p-2,p-2}(\tau).
\end{align*}
  \end{proposition}
  
  Notice that identical notation as in Proposition
  \ref{PropSolutionsEMcase} has been used despite that the fundamental
  matrices are different. Whether the solutions given in
  Propositions\ref{PropSolutionsEMcase} or
  \ref{PropSolutionsSpin2} are being referred to, should be clear
  from the context. As in the discussion of the electromagnetic case
  the components of the matrix $X^{-1}_{p,l}(0)$ will be represented
  as
  \[
    X^{-1}_{p,l}(0) =
    \begin{pmatrix}
      X^{-1}_A & X^{-1}_B \\ X^{-1}_B & X^{-1}_A
      \end{pmatrix},
  \]
  As before, the explicit form of $X^{-1}_A$ and $X^{-1}_B$ can be
  obtained inverting the fundamental matrix and by direct evaluation of
  the Jacobi polynomials. The only relevant observation is to be made
  here is that $X^{-1}_A \neq X^{-1}_B$. Expanding
  the matrix expression given in Proposition \ref{PropSolutionsSpin2}
  one has the following:
  \begin{lemma}\label{Spin2SolutionsExplicit}
    The solutions of Proposition \ref{PropSolutionsSpin2} can be
    written as
\begin{align*}
  a_{0,p;l,m}(\tau) & = C_{p,\ell,m}Q^1_{p,\ell}(\tau) +
  (-1)^{\ell}D_{p,\ell,m}Q^3_{p,\ell}(\tau), \\ a_{4,p;l,m}(\tau)& =
  D_{p,\ell,m}Q^1_{p,\ell}(-\tau) +
  (-1)^{\ell}C_{p,\ell,m}Q^3_{p,\ell}(-\tau),
\end{align*}
\begin{align*}
 & where& C_{p,\ell,m} \equiv & X^{-1}_Aa_{0,p;l,m}(0) + X^{-1}_B
  a_{4,p;l,m}(0), & D_{p,\ell,m} \equiv & X^{-1}_Ba_{0,p;l,m}(0) +
  X^{-1}_A a_{4,p;l,m}(0).
\end{align*}
  \end{lemma}  
 The $p=l$ is a special case for which one has the following:
 
   \begin{proposition}\label{Spin2LogarithmicTerms}
     For  $p\geq 2$, $\ell=p$, $-p \leq m \leq p$ one has:
\begin{align}
a_{0,p;p,m}(\tau)= \left(\frac{1-\tau}{2}\right)^{p+2}
\left(\frac{1+\tau}{2}\right)^{p-2} \left( E_{p,m}+
E^{\ast}_{p,m}\int_{0}^{\tau}\frac{\mbox{ds}}{(1+s)^{p-1}(1-s)^{p+3}}
\right),\\ a_{4,p;p,m}(\tau)= \left(\frac{1+\tau}{2}\right)^{p+2}
\left(\frac{1-\tau}{2}\right)^{p-2} \left( I_{p,m}+
I^{\ast}_{p,m}\int_{0}^{\tau}\frac{\mbox{ds}}{(1-s)^{p-1}(1+s)^{p+3}}
\right).
\end{align}
where $E_{p,l,m}$, $E^{\ast}_{p,l,m}$ and $I_{p,l,m}$,
$I^{\ast}_{p,l,m}$ are integration constants.
\end{proposition}

\begin{remark}
{\em Notice that for non-vanishing $E^{\ast}_{p,m}$ and
  $I^{\ast}_{p,m}$ the solutions
  $a_{0,p;p,m}(\tau)$ and $a_{4,p;p,m}(\tau)$ contain terms which diverge
  logarithmically near $\tau = \pm 1$.
}
\end{remark}

\subsection{Initial data for the spin-2 equations}
\label{InitialSpin2Eqns}

Consistent with equations \eqref{ExpansionGravPhin}
one considers on the initial hypersurface
 $\mathcal{S}$ fields $\phi_{n}|_{\mathcal{S}}$, 
with $n=0,1,2,3,4$ which can be expanded as
\begin{equation}
\phi_{n}|_{\mathcal{S}}= \sum_{p=|2-n|}^{\infty}\sum_{\ell=|2-n|}^{p}\sum_{m=-\ell}^{\ell}
  \frac{1}{p!}a_{n,p;\ell,m}(0)Y_{2-n;\ell m}
 \rho^{p}. \label{InitialDataExpansionGravPhin}
\end{equation}
Observe that, by virtue of equation \eqref{AlgebraicSysGrav},
 the initial data  $a_{1,p;\ell,m}(0)$, $a_{2,p;\ell,m}(0)$ and $a_{3,p;\ell,m}(0)$
is determined by  $a_{0,p,\ell,m}(0)$ and $a_{4,p,\ell,m}(0)$.
In addition, notice that, 
equations \eqref{CoupledFirstOrderGrav0}-\eqref{CoupledFirstOrderGrav4}
 are first order  while
equations \eqref{SecondOrderGrav0}-\eqref{SecondOrderGrav4} 
 are second order.  Therefore, the 
initial data $\dot{a}_{0,p,\ell,m}(0)$ and $\dot{a}_{4,p,\ell,m}(0)$ is determined,
as a consequence of equations 
\eqref{CoupledFirstOrderGrav0}-\eqref{CoupledFirstOrderGrav4}
restricted to $\mathcal{S}$, by the initial data 
$a_{0,p,\ell,m}(0)$ and $a_{4,p,\ell,m}(0)$. The latter considerations are succinctly incorporated
in the fundamental matrix.

\begin{remark}\label{remarkGeneralSolutionToTheConstraintsSpin2}
  \em{
    {Similar to the electromagnetic case, the identification of the
      coefficients 
$a_{0,p,\ell,m}(0)$ and $a_{4,p,\ell,m}(0)$ as the freely specifiable
data is enough for the purposes of this article. However, the general
parametrisation of the solutions to the constraint equations
\eqref{ConstraintGravEq5}-\eqref{ConstraintGravEq7} can be given in
terms of Hertz potentials exploiting the results of \cite{AndBacJodi14}. The general solution to the constraint equation
      \eqref{ConstraintGravEq5}-\eqref{ConstraintGravEq7} can be
      written as
\[\phi_{ABCD}= (\mathcal{G}_{4}\varphi)_{ABCD},\]
where $\varphi_{ABCD}=\varphi_{(ABCD)}$ is totally symmetric but
otherwise arbitrary spinor encoding the freely specifiable data and
$\mathcal{G}_{4}$ is an operator built using the Levi-Civita
connection $D$ associated with metric $\bmh$ intrinsic to the
hypersurface $\mathcal{S}$. Since the detailed form of the \emph{fundamental operator}
$\mathcal{G}_{4}$ and the construction of initial data using this
approach is not required for the main discussion of this article, this
will be presented elsewhere.}}
\end{remark}

\begin{assumption}\label{RegularityConditionSpin2}
  \em{ Although general initial data allows for solutions with
    $E^\star_{p,m} \neq 0$ and $I^\star_{p,m} \neq 0$ for the
    calculation of the NP constants it will be assumed that}
   \[
E^\ast_{p,m}=I^\ast_{p,m} =0.
  \]
  In other words, it will be assumed that the fields $\phi_{n}$ do not
  contain the diverging terms of Proposition
  \ref{Spin2LogarithmicTerms}.  
\end{assumption}

\begin{remark}\label{TimeSymmetricInitialDataSpin2MagneticPartVanishes}
  \em{ The spinor $\phi_{ABCD}$ can be decomposed into its electric and
    magnetic parts, denoted as $\eta_{ABCD}$ and $\mu_{ABCD}$, as
    follows:
  \[
    \phi_{AB} = \eta_{ABCD}+ \mbox{i}\mu_{ABCD},
  \]
  with
  \[
\eta_{ABCD} = \frac{1}{2}(\phi_{ABCD} + \phi^{\dagger}_{ABCD}), \qquad
\mu_{ABCD}=-\frac{1}{2}\mbox{i} \;(\phi_{ABCD} -
\phi^{\dagger}_{ABCD}),
  \]
 where $\phi^{\dagger}_{ABCD}
  \equiv
  \tau_{A}{}^{A'}\tau_{B}{}^{B'}\tau_{C}{}^{C'}\tau_{D}{}^{D'}\bar{\phi}_{A'B'C'D'}$
  with $\tau^{AA'}$ as defined in Remark
  \ref{TimeSymmetricInitialDataMaxwellMagneticPartVanishes} ---see
  \cite{CFEBook} for further discussion on the space spinor formalism.
  Initial data for which $\mu_{ABCD}|_{\mathcal{S}}=0$ will be called
  time-symmetric. A calculation then shows that time-symmetric data
  satisfy,
  \[
\phi_0 = \bar{\phi}_4, \qquad \phi_1 = -\bar{\phi}_3, \quad \phi_{2}
=\bar{\phi}_{2} \qquad \text{on} \quad \mathcal{S}
  \]
  The latter conditions imply that
  $a_{0,p,\ell,m}(0)=a_{4,p,\ell,m}(0)$.}
\end{remark}
  
  \begin{remark}\label{TimeSymmetricInitialDataSpin2}
    { \em If the data is time-symmetric then
      $C_{p,\ell,m}=D_{p,\ell,m}$. Nevertheless, for generic initial
      data one has $C_{p,\ell,m} \neq D_{p,\ell,m}$.}
  \end{remark}

\begin{remark}
{\em The convergence of the expansions \eqref{ExpansionGravPhin}
 follows from the results given in \cite{Val03a}.}
\end{remark}

\section{The NP-gauge }
\label{sec:NPGauge}

In this section, an adapted frame satisfying the
NP-gauge conditions and Bondi coordinates are constructed for
the conformal extension  introduced in Section 
\ref{TheCylinderAtSpatialInfinity}.
 For convenience of the reader, a general discussion
of the NP-gauge conditions and the construction of
 Bondi coordinates is provided
in the first part of this section.

\subsection{The NP-gauge conditions and Bondi coordinates}
\label{NPconditions}

After a brief description of the construction
the NP-frame in general asymptotically simple spacetimes,
the discussion is
particularised to the case of the Minkowski spacetime.
A more comprehensive discussion of the
NP gauge conditions in general asymptotically simple spacetimes
 can be found in \cite{Ste91,FriKan00,CFEBook}.

\medskip
Let $(\mathcal{M},\bmg,\Xi)$ denote a conformal extension of an
asymptotically simple spacetime $(\tilde{\mathcal{M}},\tilde{\bmg})$
where $\tilde{\bmg}$ satisfies the vacuum Einstein field equations
with vanishing Cosmological constant.  It is a general result in the
theory of asymptotics that for vacuum spacetimes with vanishing
Cosmological constant the conformal boundary $\mathscr{I}$, with locus
given by $\Xi=0$, consists of two disjoint null hypersurfaces
$\mathscr{I}^{+}$ and $\mathscr{I}^{-}$ each one having the topology
of $\mathbb{R}\times \mathbb{S}^2$---see \cite{Ste91,CFEBook}.  In
this section the discussion will be particularised to
$\mathscr{I}^{+}$.  Nevertheless, similar results and constructions
can be formulated, \emph{mutatis mutandi}, for $\mathscr{I}^{-}$.  To
simplify the notation, the symbol $\simeq$ will be used to denote
equality at $\mathscr{I}$, e.g. if $w$ is a scalar field on
$\mathcal{M}$ that vanishes at $\mathscr{I}$ one writes $w \simeq 0$.
Let $\{\bme'_{\bmA \bmA'} \}$ denote a frame satisfying
$\bmg(\bme'_{\bmA \bmA'},\bme'_{\bmB \bmB'}) =\epsilon_{\bmA
  \bmB}\epsilon_{\bmA' \bmB'}$ in a neighbourhood $\mathcal{U}\subset
\mathcal{M}$ of $\mathscr{I}^{+}$.  Additionally, let $\Gamma'_{\bmA
  \bmA'}{}^{\bmB}{}_{\bmC}$ denote the reduced connection coefficients
of the Levi-Civita connection of $\bmg$ defined respect to
$\bme'_{\bmA \bmA'}$.  The frame $\bme'_{\bmA \bmA'}$ is an adapted
frame at $\mathscr{I}^{+}$ if the following conditions hold:
\begin{itemize}
\item[(i)] The vector $\bme'_{\bm1 \bm1'}$ is tangent to and parallely
  propagated along $\mathscr{I}^{+}$, i.e.,
 \begin{equation*}
 \nabla_{\bm1 \bm1'} \bme'_{\bm1 \bm1'} \simeq 0.
 \end{equation*}
\item[(ii)] On $\mathcal{U}$ there exists a smooth function $u$
  inducing an affine parameter on the null generators of
  $\mathscr{I}^{+}$, namely $\bme'_{\bm1 \bm1'}(u) \simeq 1$. The
  vector $\bme'_{\bm0 \bm0'}$ is then defined as $\bme'_{\bm0
    \bm0'}=\bmg(\mbox{\textbf{d}}u, \cdot )$ so that it is tangent to
  the null generators of the hypersurfaces transverse to $\mathscr{I}$
  defined by
\begin{equation*}
\mathcal{N}_{u_{\circ}}\equiv \{ p \in \mathcal{U} \;| \;
u(p)=u_{\circ}\},
\end{equation*}
 with constant $u_{\circ}$.
\item[(iii)] The frame $\{\bme'_{\bmA \bmA'}\}$ is tangent to the cuts
  $\mathcal{C}_{u_{\circ}}\equiv \mathcal{N}_{u_{\circ}}\cap
  \mathscr{I}^{+} \approx \mathbb{S}^2$ and parallely propagated along
  $\mathcal{N}_{u_{\circ}}$, namely
\begin{equation*}
\nabla_{\bm0\bm0'}\bme'_{\bmA\bmA'}=0 \qquad \text{on} \qquad
\mathcal{N}_{u_{\circ}}.
\end{equation*}
\end{itemize}

Starting from conditions (i)-(iii) above,  in the remaining of this
section the relation between the NP and F-frame in the Minkowski
spacetime will be established. This discussion follows closely that of
\cite{FriKan00}.  The starting point is to set
\begin{equation}\label{StartingPointNP}
  \bme'{}^{a}_{\bm1\bm1'} \simeq fg^{ab}\nabla_b\Theta.
\end{equation}
A calculation using equation \eqref{MetricAndConformalFactor} renders
\begin{align*}
  g^{ab}\nabla_b\Theta = -\rho\tau(1-\tau^2)\bm\partial_\tau^a
  -\rho^2(1+\tau^2)\bm\partial_\rho^a.
\end{align*}

Thus one has $\bme'^{a}_{\bm1\bm1'} \simeq -f
\rho^2\bm\partial_\rho^a$.  Equation \eqref{StartingPointNP} and
condition (i) imply the equation 
\[
g^{ac}(f\nabla_b\nabla_c\Theta +
\nabla_bf\nabla_c\Theta)\bme'{}^{b}_{\bm1\bm1'} \simeq 0.
\]
One way to solve this equation is to consider a vector $z^a$ such that
$\nabla_z\Theta \slashed{\simeq} 0$. Transvecting the last equation
with $z_a$ and rearranging renders
\begin{equation}\label{fEquationGeneral}
g^{ab}\nabla_a \ln f \nabla_b\Theta
+\frac{\nabla_{z}g^{ab}\nabla_a\Theta\nabla_b\Theta}{2\nabla_z \Theta}
\simeq 0.
\end{equation}
Observe that $\bm\partial_\tau \Theta =-2\rho\tau$ hence $z=
\bmpartial_\tau$ is admissible. Taking this $z$, assuming that
$f=f(\rho)$ and replacing $\simeq$ with $=$ in equation
\eqref{fEquationGeneral} a direct calculation gives,
\[
 \frac{\mbox{d}\ln f}{\mbox{d}\rho} =
 -\frac{(1-\tau^2)}{\rho(1+\tau^2)}.
\]
Although the last equation could be readily solved, one is only
interested in evaluating this expression at $\mathscr{I}^{+}$ namely,
the condition is $\mbox{d}\ln f/\mbox{d}\rho \simeq 0$.  To adhere
  to the conventions of \cite{FriKan00} one sets $f=f_{\star}$, with
  $f_\star=-1/2\sqrt{2}$, which trivially solves the latter equation
  at $\mathscr{I}^{+}$. Thus, overall, one has the following condition
\begin{equation}\label{BoundaryConditionLegnNPTetrad}
\bme'_{\bm1\bm1'} \simeq -f_\star\rho^2\bm\partial_\rho.
\end{equation}
It is worth stressing that the above condition only holds at
$\mathscr{I}^+$ and that the leg $\bme'_{\bm1\bm1'}$ of the NP-frame
has not been determined on the interior of the spacetime yet.
Now, for imposing condition (ii) it is necessary for find a coordinate
$u$ such that $\bme'_{\bm1\bm1'}(u) \simeq 1$, this is
$\frac{1}{2\sqrt{2}}\rho^2\bm\partial_\rho u \simeq 1$. Direct
integration yields
\begin{equation}\label{conditionAtScriu}
  u \simeq -\frac{\sqrt{2}}{\rho} + u_\star.
\end{equation}
Again notice that this is an equality at $\mathscr{I}^+$ satisfying
that $u \rightarrow -\infty$ as $\rho \rightarrow 0$.  To determine
$u$ in the interior of the spacetime one needs to solve the \emph{eikonal
equation}
\[
g^{ab}\nabla_a\nabla_bu=0.
\]
Writing the metric in terms of the F-frame $\bme_{\bmA\bmA'}$ namely,
$g^{ab}=\epsilon^{\bmA\bmB}\epsilon^{\bmA'\bmB'}\bme^a_{\bmA\bmA'}\bme^b_{\bmB\bmB'}$,
the eikonal equation reads
\[
  \bme^a_{\bm0\bm0'}\bme^b_{\bm1\bm1'}\nabla_au\nabla_bu-\bme^a_{\bm0\bm1'}
  \bme^b_{\bm1\bm0'}\nabla_au\nabla_bu=0.
\]
Recall that the F-frame $\bme^{a}_{\bmA\bmA'}$, unlike the NP-frame
$\bme'^{a}_{\bmA\bmA'}$, is known on the interior of the spacetime
---see equation \eqref{Fframe}--- and hence the above equation can
be explicitly solved.  Assuming that $u=u(\tau,\rho)$ then one can
instead look for the solution to the simpler equation
\[
\bme_{\bm1\bm1'}^a\nabla_au=0.
\]
The latter equation can be solved using the method of characteristics and the
boundary condition \eqref{conditionAtScriu}. A direct calculation
renders
\[
u = u_\star - \frac{2\sqrt{2}}{\rho(1+\tau)}.
\]
With this expression at hand one is now in position to determine the
$\bm0\bm0'$ leg of the NP-frame in interior of the spacetime using
that
\[
  \bme'^{a}_{\bm0\bm0'} = g^{ab}\nabla_b u.
\]
A straightforward calculation then gives
\begin{equation}\label{llegNPtonlegF}
  \bme'^{a}_{\bm0\bm0'} = \frac{4}{\rho(1+\tau)^2}\bme^{a}_{\bm1\bm1'}.
\end{equation}
With the latter result one can readily check that the angular sector
of the F-frame is already parallely propagated along
$\bme'^{a}_{\bm0\bm0'}$. In other words
\[
  \bme'^{b}_{\bm0\bm0'}\nabla_b\bme^a_{\bm0\bm1'}=\bme'^{b}_{\bm0\bm0'}\nabla_b\bme^a_{\bm1\bm0'}=0.
\]
Hence, condition (iii) is automatically satisfied.  Thus, one can set
\begin{equation}\label{mmbarlegNPtombarmlegF}
  \bme'^{a}_{\bm0\bm1'} = e^{-2\mbox{i}\omega}\bme^{a}_{\bm1\bm0'},
  \qquad \bme'^{a}_{\bm1\bm0'} =
  e^{-2\mbox{i}\omega}\bme^{a}_{\bm0\bm1'}.
\end{equation}

where $\omega$ is a real number encoding a general spin rotation of the frame.
It only remains to determine $\bme'^{a}_{\bm1\bm1'}$, to do so one
could parallely propagate $\bme'^{a}_{11'}|_{\mathscr{I}^+}$ along
$\bme'^{a}_{\bm0\bm0'}$ namely, solve
\[
  \bme'^{b}_{\bm0\bm0'}\nabla_b\bme^a_{\bm1\bm1'}=0, \quad \text{with}
  \quad \bme^a_{\bm1\bm1'}\simeq -f_\star\rho^2\bm\partial_\rho^a.
\]
An alternative computationally simpler approach is to solve for
$\bme'^a_{\bm1\bm1'}$ algebraically using that
$g^{ab}=\epsilon^{\bmA\bmB}\epsilon^{\bmA'\bmB'}\bme'^a_{\bmA\bmA'}\bme'^b_{\bmB\bmB'}$
and exploiting that $\bme'^{a}_{\bm0\bm0'}$, $\bme'^{a}_{\bm0\bm1'}$
and $\bme'^{a}_{\bm1\bm0'}$ had already been determined. A calculation
using the latter approach renders
\begin{equation}\label{nlegNPtollegF}
  \bme'^{a}_{\bm1\bm1'} =
  \frac{1}{4}\rho(1+\tau)^2\bme^{a}_{\bm0\bm0'}.
\end{equation}

In the discussion of the NP-gauge, is customary to complete the
construction introducing \emph{Bondi coordinates} $(u,r)$.  Recall
that the coordinate $u$ has already been obtained.  The radial Bondi
coordinate $r$ is determined by the condition
\[
\bme'^{a}_{\bm0\bm0'}\nabla_ar=r_{\star},
\]
where $r_\star$ is a normalisation constant.  A direct calculation
using equation \eqref{llegNPtonlegF} and direct integration renders
\[
r= \frac{r_\star\rho(\tau^2-1)}{2\sqrt{2}}.
\]
Rewriting the last expression in terms of the physical coordinates
exploiting equation \eqref{UnphysicalToPhysicaltrho} and $t=\rho\tau$,
one gets
\[
r = \frac{r_\star}{2\sqrt{2}\tilde{\rho}}.
\]
To get cleaner expressions with the current normalisation of the
frame, in the remaining $r_\star=2\sqrt{2}$ will be set so that
$r=1/\tilde{\rho}$.  Therefore, the $\bm0\bm0'$ leg of the NP-frame can be
written in terms of the Bondi coordinate as
\begin{equation}\label{llegToBondiAndPhysicalRadial}
\bme'_{\bm0\bm0'}=\bm\partial_r=-\tilde{\rho}^2\bm\partial_{\tilde{\rho}}.
\end{equation}
To put the calculations of this section in the more general context of
asymptotically simple spacetimes, recall that the F-frame and the
NP-frame do not coincide because while the former is based on a Cauchy
hypersurface, the latter is adapted to $\mathscr{I}^{+}$.  However,
these frames are null frames respect to some metric $\bmg$ and
$\bmg'$, respectively, where $\bmg'=\kappa^2\bmg$, for some conformal
factor $\kappa$. Therefore, the frames $\bme_{\bmA \bmA'}$ and
$\bme'_{\bmA \bmA'}$ are, in general, related through a conformal
rescaling and a Lorentz transformation
\begin{equation}
\label{RelatingNPtoFgauge}
\bme'_{\bmA\bmA'}=\kappa^{-1}\Lambda^{\bmB}{}_{\bmA}
\bar{\Lambda}^{\bmB'}{}_{\bmA'}\bme_{\bmB \bmB'}.
\end{equation}
For the case of the Minkowski spacetime the conformal factor $\kappa$
and the Lorentz transformation $\Lambda^{\bmA}{}_{\bmB}$ can be
directly read from equations \eqref{llegNPtonlegF},
\eqref{mmbarlegNPtombarmlegF} and \eqref{nlegNPtollegF}. Furthermore,
direct inspection the frames reveals that the metric $\bmg'$
associated to the NP-frame, is related to the physical Minkowski
metric $\tilde{\bm\eta}$ via
\begin{equation}\label{gprimeTophysicalMinkowski}
  \bmg' = \frac{1}{\tilde{\rho}^2}\tilde{\bm\eta}.
\end{equation}

The discussion of this subsection can be summarised in the following:

 \begin{proposition}\label{PropNPtoFgauge}
The NP and F-frames are null frames respect to the metrics $\bmg'$ and
$\bmg$ with $\bmg'=\kappa^2\bmg$ and related via
\begin{equation}
\label{RelatingNPtoFgauge}
\bme'_{\bmA\bmA'}=\kappa^{-1}\Lambda^{\bmB}{}_{\bmA}
\bar{\Lambda}^{\bmB'}{}_{\bmA'}\bme_{\bmB \bmB'}.
\end{equation}
In the case of the Minkowski spacetime in the horizontal
representation of the cylinder at spatial infinity, the metric $\bmg$
is given by the line element \eqref{ExplicitFormMetricCylinder} and
$\bmg'= \tilde{\rho}^{-2}\tilde{\bm\eta}$ where
$\tilde{\bm\eta}$ is the physical Minkowski metric as written in equation
\eqref{MinkowskiMetricPhysicalPolar}.  Moreover, the NP-frame hinged
at $\mathscr{I}^+$ is related to the F-frame via,
\begin{align*}
 \bme'_{\bm0\bm0'}&=\frac{4}{\rho(1+\tau)^2}\bme_{\bm1\bm1'}, &
 \bme'_{\bm1\bm1'}& =\frac{1}{4}\rho(1+\tau)^2\bme_{\bm0
   \bm0'},\\ \bme'_{\bm0\bm1'}& =e^{-2 \mbox{i}\omega} \bme_{\bm1
   \bm0'}, & \bme'_{\bm1\bm0'} & = e^{-2 \mbox{i}\omega}\bme_{\bm0
   \bm1'}.
\end{align*}
The Lorentz transformation and conformal factor $\kappa$ relating the
frames is given by
\begin{equation}
\label{LorentzTransformationExplicit}
\Lambda^{\bm1}{}_{\bm0}=\frac{2
  e^{\mbox{i}\omega}}{\sqrt{\rho}(1+\tau)},\qquad
\Lambda^{\bm0}{}_{\bm1}= \frac{
  e^{-\mbox{i}\omega}\sqrt{\rho}(1+\tau)}{2}, \qquad
\Lambda^{\bm0}{}_{\bm0}=\Lambda^{\bm1}{}_{\bm1}=0, \qquad
\kappa=1.
\end{equation}
 \end{proposition}
 
Observe that the latter expressions coincide to leading order with
those reported in \cite{FriKan00} ---see also Proposition 3 of
\cite{GasVal17a}. The remaining freedom encoded in $\omega$ will be
fixed by setting $\omega=\pi$ to align with the conventions of
\cite{FriKan00}.

\section{The electromagnetic NP constants }
\label{TheElectromagneticNPConstants}

Consider the Minkowski spacetime
$(\tilde{\mathcal{M}},\tilde{\bmeta})$ as defined in Section
\ref{TheCylinderAtSpatialInfinity} and introduce the physical
retarded time $\tilde{u}=\tilde{t}-\tilde{\rho}$.  In these coordinates one has
\[
\tilde{\bmeta}=\mathbf{d}\tilde{u}\otimes \mathbf{d}\tilde{u} +
\mathbf{d}\tilde{u}\otimes\mathbf{d}\tilde{\rho} +
\mathbf{d}\tilde{\rho} \otimes
\mathbf{d}\tilde{u}-\tilde{\rho}^2\bmsigma.
\]
and the metric $\bmg'$ as given in equation \eqref{gprimeTophysicalMinkowski}.
 Let $\epsilon^{\prime}{}_{\bmA}{}^{A}$, with 
$\epsilon^{\prime}{}_{\bm0}{}^{A}=o^{\prime}{}^{A}$
and $\epsilon^{\prime}{}_{\bm1}{}^{A}=\iota'^{A}$, denote a spin dyad so that
 $e^{\prime}{}_{\bmA\bmA'}{}^{AA'}=\epsilon^{\prime}{}_{\bmA}{}^{A}
\epsilon^{\prime}{}_{\bmA'}{}^{A'}$
 constitutes the NP-frame given
in Proposition \ref{PropNPtoFgauge}.
 Let $\{\tilde{o}^{A},\tilde{\iota}^{A}\}$
 denote a spin dyad denoted by
 $\tilde{\epsilon}_{\bmA}{}^{A}$ and defined via
\begin{equation}
\label{RelationBetweenSpinDyads}
o^{A}=\tilde{\rho}\tilde{o}^A, \qquad \iota^{A}=\tilde{\iota}^A.
\end{equation}
Notice that, by virtue of  equation \eqref{gprimeTophysicalMinkowski}, the 
spin dyad   $\tilde{\epsilon}_{\bmA}{}^{A}$
 is  normalised respect to $\tilde{\bmeta}$.
 To introduce the electromagnetic NP constants as defined in
 \cite{NewPen68} consider the physical Maxwell spinor $\tilde{\phi}_{AB}$
satisfying 
\[
\tilde{\nabla}_{A'}{}^{A}\tilde{\phi}_{AB}=0,
\]
where $\tilde{\nabla}_{AA'}$ denotes the  Levi-Civita connection
respect to $\tilde{\bmeta}$.
The  components the physical Maxwell spinor respect to
 the spin dyad $\tilde{\epsilon}_{\bmA}{}^{A}$ will be denoted, as usual,
 by   $\tilde{\phi}_{0}\equiv\tilde{\phi}_{AB}\tilde{o}^{A}\tilde{o}^B$,
 $\tilde{\phi}_{1}\equiv \tilde{\phi}_{AB}\tilde{o}^{A}\tilde{\iota}^B$,
 $\tilde{\phi}_{2}\equiv
 \tilde{\phi}_{AB}\tilde{\iota}^{A}\tilde{\iota}^B$. 

\begin{assumption}
{\em Following \cite{NewPen68},  the $\tilde{\phi}_{0}$ component is
assumed to have an expansion
\begin{equation}
\label{PeelingPhysicalMaxwell}
\tilde{\phi}_{0}=\sum_{n=0}^{N}\frac{\tilde{\phi}^{n}_{0}}{\tilde{\rho}^{3+n}}
+ \mathcal{O}\Big(\frac{1}{\tilde{\rho}^{3+N}}\Big),
\end{equation}
where the coefficients $\tilde{\phi}_{0}^{n}$ do not depend on 
$\tilde{\rho}$.  }
\end{assumption}

The electromagnetic NP constants are defined through the following integrals
over cuts $\mathcal{C}$ of null infinity:
\begin{equation*}
F_{m}^{n,k}\equiv \int_{\mathcal{C}} \bar{Y}_{1;n+1,m}\tilde{\phi}_{0}^{n+1}\mbox{d}S,
\end{equation*}
where $n,\,m \in \mathbb{Z}$ with $n \geq 0$,
$|m|\leq n+1$ and $\mbox{d}S$ denotes the area element 
respect to $\bm\sigma$.
In flat space, $F_{m}^{n}$ are absolutely conserved
in the sense that their value is independent of the cut
$\mathcal{C}$ on which they are evaluated ---see \cite{NewPen68}.
From these, only those given by $n=0$ and $m=-1,0,1$ are conserved
in the general non-linear Einstein Maxwell theory ---see
\cite{NewPen68}.

\subsection{Translation to the F-gauge}
\label{TranslationToTheFGauge}
In view of equation \eqref{gprimeTophysicalMinkowski}, one has that,
as a consequence of the standard conformal transformation law
for the spin-1 equation ---see \cite{Ste91},
   the spinor $\phi'_{AB}$, satisfying
\[
\nabla'_{A'}{}^{A}\phi'_{AB}=0,
\] where $\nabla'_{AA'}$ is the Levi-Civita connection of $\bmg'$,
 is related to  $\tilde{\phi}_{AB}$ via
\begin{equation}
\label{UnphysicalMaxwellAndPhysicalMaxwellSpinor}
\phi'_{AB}=\tilde{\rho}\tilde{\phi}_{AB}.
\end{equation}
Therefore, using equations
 \eqref{PeelingPhysicalMaxwell}, \eqref{RelationBetweenSpinDyads}
and  \eqref{UnphysicalMaxwellAndPhysicalMaxwellSpinor}, 
 one obtains
\begin{equation*}
\phi'_{0}=\sum_{n=0}^{N}\frac{\tilde{\phi}^{n}_{0}}{\tilde{\rho}^n} 
+ \mathcal{O}\Big(\frac{1}{\tilde{\rho}^{N}}\Big),
\end{equation*}
where $\phi'_{0}\equiv\phi'_{AB}o'^{A}o'^{B}$. Applying $\bme'_{\bm0\bm0'}$ to the last expression and
  using equation \eqref{llegToBondiAndPhysicalRadial} one gets
\begin{equation*}
\bme'_{\bm0\bm0'}(\phi'_{0})=
\tilde{\phi}^{1}_{0} + \mathcal{O}(\tilde{\rho}^{-1}).
\end{equation*}
The repeated application of $\bme'_{\bm0\bm0'}$ to the above relation
shows that in general
\begin{equation*}
\bme'^{(q)}_{\bm0\bm0'}(\phi'_{0})=
q!\;\tilde{\phi}^{q}_{0} + \sum_{i=q+1}^{N}
\frac{(i+1)!}{(i-q+1)!}\; \frac{\tilde{\phi}^{i}_{0}}{\tilde{\rho}^{i-q}}
+ \mathcal{O}\Big(\frac{1}{\tilde{\rho}^{N-q}}\Big),
\end{equation*}
 where $\bme^{(q)}_{\bm0\bm0'}(\phi'_{0})$
denotes $q$ consecutive applications of $\bme'_{\bm0\bm0'}$ to
$\phi_{0}$. Thus, the quantities $F_{m}^{n}$ can be written as
\begin{equation}
\label{EMNPquantitiesInNPgauge}
F_{m}^{n}= \frac{1}{(n+1)! }
\frac{}{}\int_{\mathcal{C}}\bar{Y}_{1;n+1,m}\;\bme'^{(n+1)}_{\bm0\bm0'}(\phi'_{0})\;\mbox{d}S.
\end{equation}
Observe that the constants $F_{m}^{n}$ in the previous equation are
expressed in terms of $\bmg'$-associated quantities. In order to
obtain a general expression for the electromagnetic NP quantities in
the F-gauge one has to rewrite expression
\eqref{EMNPquantitiesInNPgauge} in terms of $\bmg$-related
quantities.  As discussed before, the frames $\bme_{\bmA\bmA'}$ and
$\bme'_{\bmA \bmA'}$ are related through a conformal rescaling and a
Lorentz transformation as given in equation \eqref{RelatingNPtoFgauge}. For
the sake of generality, the first part of the
 discussion will be carried out for general
 $\kappa$ and $\Lambda^{\bmA}{}_{\bmB}$.

\subsubsection{Explicit computation of the first three constants}
\label{ExplicitCompFirstThreeConstants}

Let $\epsilon_{\bmA}{}^{A}$, with $\epsilon_{\bm0}{}^{A}=o^{A}$ and
$\epsilon_{\bm1}{}^{A}=\iota^{A}$, denote a spin dyad normalised
respect to $\bmg$ as defined in Section
 \ref{TheMaxwellEquationsInTheFGauge}. As a consequence of equation
\eqref{RelatingNPtoFgauge}, the spin dyads
$\epsilon_{\bmA}{}^{A}$ and $\epsilon^{\prime}{}_{\bmA}{}^{A}$, giving rise to
 $\bme_{\bmA\bmA'}$ and $\bme'_{\bmA \bmA'}$, are related via
\begin{equation}\label{RelationSpinDyadsEqs}
\epsilon^{\prime}{}_{\bmA}{}^{A}=\kappa^{-1/2}\Lambda^{\bmB}{}_{\bmA}\epsilon_{\bmB}{}^{A}.
\end{equation}
Additionally, the spinor field $\phi_{AB}$, satisfying 
\begin{equation*}
\nabla_{A'}{}^{A}\phi_{AB}=0,
\end{equation*}
where $\nabla_{AA'}$ is the Levi-Civita connection respect to $\bmg$,
is related to $\phi'_{AB}$ via
\begin{equation*}
\phi'_{AB}=\kappa^{-1}\phi_{AB}.
\end{equation*}
Therefore, one has that
\begin{equation}\label{PhiNPMaxwellToPhiFMaxwell}
\phi'_{0}=\kappa^{-2}\Lambda^{\bmC}{}_{\bm0}\Lambda^{\bmD}{}_{\bm0}\phi_{\bmC \bmD},
\end{equation}
where $\phi_{\bmC\bmD}\equiv\epsilon_{\bmC}{}^{C}\epsilon_{\bmD}{}^{D}\phi_{CD}$.
Using the Leibniz rule one obtains
\begin{equation}
\label{LeibnizRuleAppliedOnce}
\bme'_{\bm0\bm0}(\phi'_{0})=  \kappa^{-2} \Big(
 \Lambda^{\bmC}{}_{\bm0}\Lambda^{\bmD}{}_{\bm0}\bme'_{\bm0\bm0'}(\phi_{\bmC\bmD}) +
 2\phi_{\bmC\bmD}\Lambda^{\bmC}{}_{\bm0}\bme'_{\bm0\bm0'}(\Lambda^{\bmD}{}_{\bm0})
 - 2\kappa^{-1}\Lambda^{\bmC}{}_{\bm0}\Lambda^{\bmD}{}_{\bm0}\phi_{\bmC\bmD}\bme'_{\bm0\bm0'}(\kappa) \Big).
\end{equation}
Notice that, in the above expression, all the quantities except for the
frame derivative $\bme'_{\bm0\bm0}$ are $\bmg$ related quantities,
namely, given in the F-gauge and the F-coordinates.  Using
equation \eqref{RelatingNPtoFgauge} one can expand expression
\eqref{LeibnizRuleAppliedOnce}. This leads to
the following expression for the conserved quantities:
\begin{multline}
  F_{m}^{0}=\int_{\mathcal{C}}\bar{Y}_{1;1,m}\kappa^{-3} 
       \Big( \Lambda^{\bmC}{}_{\bm0}\Lambda^{\bmD}{}_{\bm0}
\Lambda^{\bmB}{}_{\bm0}\bar{\Lambda}^{\bmB'}{}_{\bm0'}\bme_{\bmB
      \bmB'}(\phi_{\bmC \bmD})  \\
 + 2\kappa
    \phi_{\bmC\bmD}\Lambda^{\bmC}{}_{\bm0}\bme'_{\bm0\bm0'}(\Lambda^{\bmD}{}_{\bm0}) 
 - 2 \Lambda^{\bmC}{}_{\bm0}\Lambda^{\bmD}{}_{\bm0}\phi_{\bmC\bmD}\bme'_{\bm0 \bm0}(\kappa)
        \Big) \mbox{d}S, \label{EMcounterpartOfEquationIII5}
\end{multline}
for $m=-1,0,1$. These correspond to the three electromagnetic NP
quantities that remain conserved in the non-linear Einstein Maxwell
theory. The last expression represent the electromagnetic counterpart
of the gravitational NP quantities in the F-gauge as reported in
\cite{FriKan00} in equation (III.5). To simplify and
prepare the notation for the calculation of higher the order constants
$F^n_m$, it is convenient to introduce the following short-hands:
\begin{equation}\label{SimplifyingNotation}
  \Lambda \equiv
  \Lambda^{\bm1}{}_{\bm{0}}=\frac{-2}{\sqrt{\rho}(1+\tau)}, \qquad \bme
  \equiv \sqrt{2} \bme_{\bm1 \bm1'} = (1+\tau)\partial_\tau
  -\rho\partial_\rho.
\end{equation}
Then, using the results of Proposition
\ref{PropNPtoFgauge}, equation
\eqref{PhiNPMaxwellToPhiFMaxwell} and the notation
introduced in equation \eqref{SimplifyingNotation} one has
\begin{equation}\label{SimplifyingNotation2}
\phi'_{0}=\Lambda^2\phi_2, \qquad \bme'_{\bm0\bm0'} =
\frac{\sqrt{2}}{2} \Lambda^2\bme.
\end{equation}
With this notation, equation
\eqref{LeibnizRuleAppliedOnce} reduces to
\begin{equation}
\label{FirstDerivativeExplicitPrevious}
\bme'_{\bm0\bm0'}(\phi_{0}')
= \frac{\sqrt{2}}{2}(\Lambda^4\bme \phi_2 + 2\Lambda^3\phi_2 \bme \Lambda) .
\end{equation}
Additionally, using equation \eqref{SimplifyingNotation}, one gets
\begin{equation}
\label{1StDerivativeLambda}
\bme \Lambda = -\tfrac{1}{2}\Lambda.
\end{equation}
Thus, together equations \eqref{1StDerivativeLambda} and
\eqref{FirstDerivativeExplicitPrevious} read
\begin{equation}
\label{FirstDerivativeExplicitMaxwell}
\bme'_{\bm0\bm0'}(\phi_{0}')=\frac{\sqrt{2}}{2}\Lambda^4(\bme \phi_2 - \phi_2).
\end{equation}
Using that
\begin{equation}\label{bmearhop}
  \bme(a_{2,p,\ell,m}(\tau)\rho^p)= ((1+\tau)\dot{a}_{2,p,l,m}(\tau)-p
  a_{2,p,\ell,m}(\tau))\rho^p,
\end{equation}
and the Ansatz \eqref{ExpansionPhin} a calculation renders
\[
  \bme'_{\bm0\bm0'}(\phi_{0}') = \frac{\sqrt{2}}{2}
  \sum_{p=1}^{\infty}\sum_{\ell=1}^{p}\sum_{m=-\ell}^{\ell}
  \frac{1}{p!}\Lambda^4 \rho^p ((1+\tau)\dot{a}_{2,p,l,m}(\tau)-(1+p)
  a_{2,p,\ell,m}(\tau))Y_{1;\ell, m}.
  \]
Separating the dependence in $\tau$ and $\rho$ explicitly, the latter
expression can be written as
\begin{equation}\label{FirstDerivativeMaxwellKeyExpression}
  \bme'_{\bm0\bm0'}(\phi_{0}') = 2^4 \Big(\frac{\sqrt{2}}{2}\Big)
  \sum_{p=1}^{\infty}\sum_{\ell=1}^{p}\sum_{m=-\ell}^{\ell}
  \frac{1}{p!}(1+\tau)^{-4}
  \rho^{(p-2)}A^{1}_{2,p,l,m}(\tau)Y_{1;\ell, m},
\end{equation}
where
\begin{align}\label{FirstOperatorMaxwell}
A^{1}_{2,p,l,m}(\tau) \equiv ((1+\tau)\dot{a}_{2,p,l,m}(\tau)-(1+p)
a_{2,p,\ell,m}(\tau)).
\end{align}
The subindices in $A^{1}_{2,p,l,m}$ are a copy of those of $a_{2,p,l,m}$
while the superindex  counts the number of times that the
$\bme'_{\bm0\bm0'}$-derivative was applied.
Let $A^{1}_{2,p,l,m}|_{\mathscr{I}^{+}}$ 
  denote $A^{1}_{2,p,l,m}(\tau=1)$.  Using equation
  \eqref{FirstDerivativeMaxwellKeyExpression} and
  \eqref{EMNPquantitiesInNPgauge} computing the first set of NP
  constants reduces to the evaluating the following
  integral
  \begin{align}
  \label{FirstConstantsSimplifiedStepThree}
F_{m}^{0}=2^4 \Big(\frac{\sqrt{2}}{2}\Big)\lim_{\substack{\rho \to \rho_\star \\ \tau \to 1}}\Bigg(
\int_{\mathbb{S}^2}
\sum_{p=1}^{\infty}\sum_{\ell=1}^{p}\sum_{m=-\ell}^{\ell}
\frac{1}{p!}(1+\tau)^{-4} \rho^{(p-2)}A^{1}_{2,p,l,m}(\tau)Y_{1;\ell,
  m}\bar{Y}_{1;1,m} \mbox{d}S \Bigg).
\end{align}
where $\rho_\star$ is a constant that parametrises the choice of cut,
$\mathcal{C}$, of $\mathscr{I}^{+}$. In particular,
$\rho_\star=0$ represents the choice $\mathcal{C}=I^{+}$. Using the
orthogonality relation,
\begin{equation}
\label{OrthogonalitySphericalHarmonics}
\int_{\mathbb{S}^2}Y_{s;\ell',m'}\bar{Y}_{s;\ell,m}=\delta_{\ell,\ell'}\delta_{m,
  m'},
\end{equation}
one obtains,
\begin{align}\label{PreFirstNPMaxwellConstant}
F_{m}^{0}=2^4 \Big(\frac{\sqrt{2}}{2}\Big) \sum_{p=1}^{\infty}
\frac{1}{p!}\rho_{\star}^{(p-2)}A^{1}_{2,p,1,m}|_{\mathscr{I}^{+}}.
\end{align}
Naively one would conclude that $F_{m}^{0}$ is singular
if the cut $I^{+}$ is chosen and that equation
\eqref{PreFirstNPMaxwellConstant} contains an infinite number of terms,
however, a direct calculation using the explicit form of
$a_{2,p,l,m}(\tau)$ shows that

\begin{remark}\label{remarkMagicCancelationFirstDerivativeMaxwell}
   $A^{1}_{2,p,1,m}|_{\mathscr{I}^{+}} = 0$ for $p=1$ and $p \geq 3$.
\end{remark}
Exploiting Remark \ref{remarkMagicCancelationFirstDerivativeMaxwell},
one concludes that
\begin{align}\label{FirstNPMaxwellConstant}
F_{m}^{0}= -\frac{3\sqrt{2}}{2} C_{2,1,m},
\end{align}
where $C_{2,1,m}$ is a constant determined by the initial data as in
Lemma \ref{MaxwellSolutionsExplicit}.  Before computing the next set
of constants in the hierarchy, $F_{m}^{1}$, a couple of observations
are in order.  The NP constant $F_{m}^{0}$ comes from the term with
$p=2$ in equation \eqref{FirstConstantsSimplifiedStepThree}. All the
terms with $p \geq 3$ and $l=1$ that could potentially contribute to
the NP constant vanish when $A^{1}_{2,p,l,m}$ is evaluated at
$\mathscr{I}^{+}$.  This is not surprising as the NP constants are
independent of the cut $\mathcal{C}$ on which they are computed
(constancy). In doing this calculation Assumption
\ref{RegularityConditionMaxwell} has been used in order to discard
the logarithmic terms of Proposition \ref{MaxwellLogarithmicTerms}.
This is a necessary restriction on the initial data to get well
defined NP constants ---see \cite{Val98} and
\cite{Val99a} for a discussion about the connection between the
Peeling theorem and the \emph{classical NP constants}
and how an alternative set of
\emph{logarithmic NP constants} can be defined in the polyhomogeneous
case. Additionally, notice that, if one selects the cut $\mathcal{C} =
I^{+}$ --- hence $\rho_\star=0$--- and invoke the \emph{finiteness and
  constancy} of the NP constants, Remark
\ref{remarkMagicCancelationFirstDerivativeMaxwell} is not needed.  In
fact, Remark \ref{remarkMagicCancelationFirstDerivativeMaxwell} can be
thought as an alternative proof of the finiteness and constancy of the
first set of electromagnetic NP constants in Minkowski spacetime.

\medskip
To compute the next set of constants in the hierarchy $F_{m}^1$
  observe that applying $\bme_{\bm0 \bm0'}'$ to equation
  \eqref{FirstDerivativeExplicitMaxwell} gives
\begin{align}\label{SecondDerivativeExplicit}
\bme'{}^{(2)}_{\bm0\bm0'}{}\phi'_{0} =\Big(\frac{\sqrt{2}}{2}\Big)^2\Lambda^{6}(\bme^{(2)}
\phi_2 -3\bme\phi_2+2 \phi_2).
\end{align}
Then, using that
\begin{equation}\label{bmebmearhop}
  \bme^{(2)}(a_{2,p,\ell,m}(\tau)\rho^p)=
  \Big((1+\tau)^2\ddot{a}_{2,p,l,m}(\tau) +(1+\tau)(1-2p)\dot{a}_{2,p,\ell,m}(\tau)
  + p^2a_{2,p,\ell,m}(\tau)\Big)\rho^p,
\end{equation}
and equation \eqref{bmearhop}, a calculation renders
\begin{equation}\label{SecondDerivativeMaxwellKeyExpression}
  \bme'{}_{\bm0\bm0'}^{(2)}(\phi_{0}') = 2^6 \Big(\frac{\sqrt{2}}{2}\Big)^2
  \sum_{p=1}^{\infty}\sum_{\ell=1}^{p}\sum_{m=-\ell}^{\ell}
  \frac{1}{p!}(1+\tau)^{-6}
  \rho^{(p-3)}A^{2}_{2,p,l,m}(\tau)Y_{1;\ell, m},
\end{equation}
where
\begin{align}\label{SecondOperatorMaxwell}
A^{2}_{2,p,l,m}(\tau) \equiv
(1+\tau)^2\ddot{a}_{2,p,l,m}(\tau)-2(1+\tau)(1+p)
\dot{a}_{2,p,\ell,m}(\tau) + (1+p)(2+p)a_{2,p,\ell,m}(\tau).
\end{align}
Thus, using equation \eqref{EMNPquantitiesInNPgauge} gives
\begin{align}
\label{SecondConstantsSimplifiedStepThree}
F_{m}^{1}= 2^6 \Big(\frac{\sqrt{2}}{2}\Big)^2 \Big(\frac{1}{2!}\Big)
\lim_{\substack{\rho \to \rho_\star \\ \tau \to
    1}}\Bigg( \int_{\mathbb{S}^2}
\sum_{p=1}^{\infty}\sum_{\ell=1}^{p}\sum_{m=-\ell}^{\ell}
\frac{1}{p!}(1+\tau)^{-6} \rho^{(p-3)}A^{2}_{2,p,l,m}(\tau)Y_{1;\ell,
  m}\bar{Y}_{1;2,m} \mbox{d}S \Bigg).
\end{align}
Exploiting the orthogonality condition
\eqref{OrthogonalitySphericalHarmonics} renders
\begin{align}\label{PreSecondNPMaxwellConstant}
  F_{m}^{1}= 2^6 \Big(\frac{\sqrt{2}}{2}\Big)^2  \Big(\frac{1}{2!}\Big)
  \sum_{p=1}^{\infty}
\frac{1}{p!}\rho_{\star}^{(p-3)}A^{2}_{2,p,2,m}|_{\mathscr{I}^{+}}.
\end{align}
A direct calculation using equation \eqref{SecondOperatorMaxwell} and
the explicit form of the solution $a_{2,p,2,m}(\tau)$ gives the following:

\begin{remark}\label{remarkMagicCancelationSecondDerivativeMaxwell}
   $A^{2}_{2,p,2,m}|_{\mathscr{I}^{+}} = 0$ for $p \leq 2$ and $p \geq 4$.
\end{remark}
Using equations \eqref{remarkMagicCancelationSecondDerivativeMaxwell} and
\eqref{PreFirstNPMaxwellConstant} one concludes that
\begin{align}\label{FirstNPMaxwellConstant}
F_{m}^{1}= \frac{20}{3} C_{3,2,m}.
\end{align}
Notice again that the constant comes from the $p=3$ term in equation
\eqref{SecondConstantsSimplifiedStepThree} and that Remark
\ref{remarkMagicCancelationSecondDerivativeMaxwell} can be avoided if
one invokes the finiteness and constancy of the NP
constants  $F_{m}^{1}$ in order to evaluate them directly at $\rho_{\star}=0$.
Before discussing the general case, it is instructive to compute the
next constant in the hierarchy: $F_{m}^{2}$. Proceeding as with the
previous set of constants, applying $\bme'_{\bm0\bm0'}$ to equation
\eqref{SecondDerivativeExplicit} gives
\begin{align}\label{ThirdDerivativeExplicit}
  \bme'^{(3)}_{\bm0\bm0'}{}\phi'_{0} =
  \Big(\frac{\sqrt{2}}{2}\Big)^3\Lambda^{8}(\bme^{(3)} \phi_2
  -6\bme^{(2)}\phi_2 + 11\bme\phi_2 -6 \phi_2).
\end{align}
A calculation using that
\begin{multline}\label{bmebmebmearhop}
  \bme^{(3)}(a_{2,p,\ell,m}(\tau)\rho^p)=
  \Big((1+\tau)^3\dddot{a}_{2,p,l,m}(\tau)
  -3(1+\tau)(p-1)\ddot{a}_{2,p,l,m}(\tau) +
  \\ (1+\tau)(1-3p+3p^2)\dot{a}_{2,p,l,m}(\tau) -p^3 a_{2,p,l,m}(\tau)
  \Big)\rho^p,
\end{multline}
renders
\begin{equation}\label{ThirdDerivativeMaxwellKeyExpression}
  \bme'{}_{\bm0\bm0'}^{(3)}(\phi_{0}') = 2^8
  \Big(\frac{\sqrt{2}}{2}\Big)^3
  \sum_{p=1}^{\infty}\sum_{\ell=1}^{p}\sum_{m=-\ell}^{\ell}
  \frac{1}{p!}(1+\tau)^{-8}
  \rho^{(p-4)}A^{3}_{2,p,l,m}(\tau)Y_{1;\ell, m},
\end{equation}
where
\begin{multline}\label{ThirdOperatorMaxwell}
A^{3}_{2,p,l,m}(\tau) \equiv
(1+\tau)^3\dddot{a}_{2,p,l,m}(\tau)-3(1+\tau)^2(1+p)\ddot{a}_{2,p,\ell,m}(\tau)
\\ + 3(1+\tau)(1+p)(2+p)\dot{a}_{2,p,\ell,m}(\tau) - (3+p)(2+p)(1+p)
a_{2,p,\ell,m}(\tau) .
\end{multline}
Then, the orthogonality condition
\eqref{OrthogonalitySphericalHarmonics} and expression
\eqref{EMNPquantitiesInNPgauge} give
\begin{align}\label{PreThirdNPMaxwellConstant}
F_{m}^{2}= 2^8 \Big(\frac{\sqrt{2}}{2}\Big)^3 \Big(\frac{1}{3!}\Big)
\sum_{p=1}^{\infty}
\frac{1}{p!}\rho_{\star}^{(p-4)}A^{3}_{2,p,3,m}|_{\mathscr{I}^{+}}.
\end{align}
To simplify the latter equation, a direct calculation using equation
\eqref{ThirdOperatorMaxwell} and the explicit form of the solution
$a_{2,p,3,m}$ gives the following:
\begin{remark}\label{remarkMagicCancelationThirdDerivativeMaxwell}
   $A^{2}_{2,p,3,m}|_{\mathscr{I}^{+}} = 0$ for $p\leq 3$ and $p \geq 5$.
\end{remark}
Using Remark \ref{remarkMagicCancelationThirdDerivativeMaxwell} one
finally obtains
\begin{align}\label{ThirdNPMaxwellConstant}
F_{m}^{2}= -\frac{35\sqrt{2}}{6} C_{4,3,m}.
\end{align}

For the calculation of $F_{m}^n$ instead of proving the generalisation
of Remark \ref{remarkMagicCancelationThirdDerivativeMaxwell}, the
calculation will be simplified invoking the finiteness and constancy of
the NP constants to compute them directly at $I^+$.

\subsubsection{The general case}
The previous discussion suggests that, in principle, it should be
possible to obtain a general formula for $F_{m}^{n}$. Revisiting the calculation
of $F_{m}^{0}$, $F_{m}^{1}$ and $F_{m}^{2}$ one can obtain the following 
results concerning the overall structure of the electromagnetic NP constants
in flat space:

\begin{lemma}\label{LemmaDerivativesExplicitGeneralStructure}
For any integer $n \geq 1$,
\begin{equation}\label{nDerivativeMaxwellKeyExpression}
  \bme'{}_{\bm0\bm0'}^{(n)}(\phi_{0}') = 2^{2(n+1)}
  \bigg(\frac{\sqrt{2}}{2}\bigg)^n
  \sum_{p=1}^{\infty}\sum_{\ell=1}^{p}\sum_{m=-\ell}^{\ell}
  \frac{1}{p!}(1+\tau)^{-2(n+1)}
  \rho^{p-(n+1)}A^{n}_{2,p,l,m}(\tau)Y_{1;\ell, m},
\end{equation}
with,
\begin{equation}\label{nOperatorMaxwell}
  A^{n}_{2,p,l,m}(\tau) \equiv \sum_{k=0}^{n}(-1)^k {n \choose
    k}\frac{(k+p)!}{p!}(1+\tau)^{n-k}a^{(n-k)}_{2,p,l,m},
\end{equation}
where $a^{(k)}_{2,p,l,m} \equiv (\bm\partial_\tau)^{k}a_{2,p,l,m}$.
\end{lemma}

\begin{proof}
  To prove this result one proceeds by induction.  Equations
  \eqref{FirstOperatorMaxwell}, \eqref{SecondOperatorMaxwell} and
  \eqref{ThirdOperatorMaxwell} already show that the result is valid
  for $n=1$, $n=2$ and $n=3$.  This constitutes the basis of
  induction.
  Now, assume that expressions \eqref{nDerivativeMaxwellKeyExpression}-\eqref{nOperatorMaxwell} hold (induction
hypothesis), then applying $\bme'_{\bm0\bm0'}$ to equation
\eqref{nDerivativeMaxwellKeyExpression}, a direct calculation exploiting
equations \eqref{SimplifyingNotation} and \eqref{SimplifyingNotation2}
renders
\begin{multline}\label{nplusoneDerivativeMaxwellKeyExpression}
  \bme'{}_{\bm0\bm0'}^{(n+1)}(\phi_{0}') = 2^{2(n+2)}
  \bigg(\frac{\sqrt{2}}{2}\bigg)^{(n+1)}
  \sum_{p=1}^{\infty}\sum_{\ell=1}^{p}\sum_{m=-\ell}^{\ell}
  \frac{1}{p!}(1+\tau)^{-2(n+2)}
  \rho^{p-(n+2)}R^{n}_{2,p,l,m}(\tau)Y_{1;\ell, m} ,
\end{multline}
where
\[
R^{n}_{2,p,l,m}(\tau) = (1+\tau)\dot{A}^{n}_{2,p,l,m}(\tau)
  - (p+n+1) A^{n}_{2,p,l,m}(\tau).
\]
Substituting \eqref{nOperatorMaxwell} into the last expression  gives
\begin{multline}
  R^{n}(\tau)= (1+\tau)\sum_{k=0}^{n}(-1)^k {n \choose
    k}\frac{(k+p)!}{p!}\Big((n-k)(1+\tau)^{n-k-1}a^{(n-k)} +
  (1+\tau)^{n-k}a^{(n-k-1)}\Big) \\ -(p+n+1)\sum_{k=0}^{n}(-1)^k {n
    \choose k}\frac{(k+p)!}{p!}(1+\tau)^{n-k}a^{(n-k)},
\end{multline}
where the subindices in $R^{n}_{2,p,\ell,m}$ and
$a_{2,p,\ell,m}$ have been omitted for conciseness.
Reorganising the terms one gets
\begin{multline}\label{RnExpressionIntermediateMaxwell}
  R^{n}(\tau)= \sum_{k=0}^{n}(-1)^k {n \choose
    k}\frac{(k+p)!}{p!}(1+\tau)^{n-k+1}a^{(n-k+1)}
  \\ -\sum_{k=0}^{n}(-1)^k {n \choose k}\frac{(k+p)!}{p!}(p+k+1)(1+\tau)^{n-k}a^{(n-k)}.
\end{multline}
Expanding the first term in the first sum of equation \eqref{RnExpressionIntermediateMaxwell} renders
\begin{multline*}
  R^{n}(\tau)= (1+\tau)^{n+1}a^{(n+1)} +
  \sum_{\lambda=0}^{n-1}(-1)^{\lambda +1} {n \choose \lambda + 1
  }\frac{(p+\lambda +1)!}{p!}(1+\tau)^{n-\lambda}a^{(n-\lambda)}
  \\ +\sum_{k=0}^{n}(-1)^{k+1} {n \choose
    k}\frac{(p+k+1)!}{p!}(1+\tau)^{n-k}a^{(n-k)}.
\end{multline*}
Separating the last term in the second sum and rearranging gives
\begin{multline*}
  R^{n}(\tau)= (1+\tau)^{n+1}a^{(n+1)} +
  \sum_{k=0}^{n-1}(-1)^{k+1}\frac{(p+k
    +1)!}{p!}(1+\tau)^{n-k}a^{(n-k)}\Bigg( {n \choose k } + {n \choose
    n+1 } \Bigg) \\ + (-1)^{n+1}\frac{(p+n+1)!}{p!} a.
\end{multline*}
Using the recursive identity of the binomial coefficients
\begin{equation}\label{BinomialRecursive}
{i \choose j } = {i-1 \choose j } + {i-1 \choose j-1 },
\end{equation}
renders
\begin{multline*}
  R^{n}(\tau)= (1+\tau)^{n+1}a^{(n+1)} +
  \sum_{k=0}^{n-1}(-1)^{k+1}{n+1 \choose k+1 }\frac{(p+k
    +1)!}{p!}(1+\tau)^{n-k}a^{(n-k)}\\ + (-1)^{n+1}\frac{(p+n+1)!}{p!}
  a.
\end{multline*}
Equivalently,
\begin{multline*}
  R^{n}_{2,p,l,m}(\tau)= (1+\tau)^{n+1}a^{(n+1)} +
  \sum_{i=0}^{n-1}(-1)^{i}{n+1 \choose i
  }\frac{(p+i)!}{p!}(1+\tau)^{n-i-1}a^{(n-i-1)}\\ +
  (-1)^{n+1}\frac{(p+n+1)!}{p!} a.
\end{multline*}
Relabelling the counter to absorb the first and last term and using
the definition given in equation \eqref{nOperatorMaxwell} one gets
\[
R^{n}_{2,p,\ell,m}(\tau)=A^{n+1}_{2,p,\ell,m}(\tau).
\]
Finally, substituting the last expression into
equation \eqref{nplusoneDerivativeMaxwellKeyExpression} one concludes that
\begin{multline*}
  \bme'{}_{\bm0\bm0'}^{(n+1)}(\phi_{0}') = 2^{2(n+2)}
  \bigg(\frac{\sqrt{2}}{2}\bigg)^{(n+1)}
  \sum_{p=1}^{\infty}\sum_{\ell=1}^{p}\sum_{m=-\ell}^{\ell}
  \frac{1}{p!}(1+\tau)^{-2(n+2)}
  \rho^{p-(n+2)}A^{n+1}_{2,p,l,m}(\tau)Y_{1;\ell, m}  ,
\end{multline*}
which completes the induction step.
\end{proof}
With Lemma \ref{LemmaDerivativesExplicitGeneralStructure} at hand,
it is straightforward to determine the general structure of the NP
constants:
\begin{proposition}\label{PropositionNPconstantsGeneralStructure}
  The NP constants in Minkowski spacetime $F^{n-1}_{m}$ are given by
  \[
    F_{m}^{n-1}= Q^+(n) C_{n+1,n,m},
\]
  where $Q^+(n)$ is a numerical factor and
  $C_{n+1,n,m}$ is determined by the initial data $a_{0,n+1,n,m}(0)$
  and $a_{2,n+1,n,m}(0)$ as given in Lemma \ref{MaxwellSolutionsExplicit}.
\end{proposition}
\begin{proof}
Exploiting the orthogonality condition
\eqref{OrthogonalitySphericalHarmonics}, the expression for the
electromagnetic NP constants given in
equation \eqref{EMNPquantitiesInNPgauge} and equations
\eqref{nDerivativeMaxwellKeyExpression} and \eqref{nOperatorMaxwell}
render
\begin{align}\label{PrenthNPMaxwellConstant}
  F_{m}^{n-1}= 2^{n+1} \bigg(\frac{\sqrt{2}}{2}\bigg)^n
  \Big(\frac{1}{n!}\Big) \sum_{p=1}^{\infty}
  \frac{1}{p!}\rho_{\star}^{(p-(n+1))}A^{n}_{2,p,n,m}|_{\mathscr{I}^{+}}.
\end{align}
Invoking the finiteness and constancy of the NP constants in order to
evaluate the last expression on the cut $I^+$ one concludes that

\begin{align}\label{nthNPMaxwellConstant}
F_{m}^{n-1}= 2^{n+1}
\bigg(\frac{\sqrt{2}}{2}\bigg)^n\bigg(\frac{1}{n!(n+1)!}\bigg)A^{n}_{2,n+1,n,m}|_{\mathscr{I}^{+}}.
\end{align}
Direct inspection of $A^{n}_{2,n+1,n,m}|_{\mathscr{I}^{+}}$
renders
\begin{align}\label{nNPMaxwellConstantInTermsofInitialData}
F_{m}^{n-1}= Q^+(n) C_{n+1,n,m},
\end{align}
where $Q^+(n)$ is an irrelevant numerical factor.
\end{proof}

\subsection{The constants at $\mathscr{I}^-$}
The analysis carried out in Sections \ref{sec:NPGauge} and
\ref{TheElectromagneticNPConstants} for the electromagnetic constants
defined at $\mathscr{I}^{+}$, can be performed in a completely
analogous way for $\mathscr{I}^{-}$.  To do so, consider a formal
replacement $\tau \rightarrow-\tau$. Upon this
formal replacement the roles of $\bm\ell = \bme_{\bm0 \bm0'}$ and
$\bmn = \bme_{\bm1\bm1'}$ as defined in \eqref{Fframe0011} and
$\phi_{0}$ and $\phi_{2}$ are essentially interchanged.  Then,
following the discussion of Sections \ref{sec:NPGauge} and
\ref{TranslationToTheFGauge} one obtains \emph{mutatis mutandis} the
time dual of Proposition \ref{PropNPtoFgauge} and
\ref{PropositionNPconstantsGeneralStructure}:

\begin{proposition}\label{PropNPtoFgaugePast}
  The NP-frame hinged at
$\mathscr{I}^-$ is related to the F-frame via
\begin{align*}
 \bme'_{\bm1\bm1'}&=\frac{4}{\rho(1-\tau)^2}\bme_{\bm0\bm0'} &
 \bme'_{\bm0\bm0'}& =\frac{1}{4}\rho(1-\tau)^2\bme_{\bm1\bm1'},\\
 \bme'_{\bm1\bm0'}& =e^{-2 \mbox{i}\omega} \bme_{\bm0 \bm1'}, &
 \bme'_{\bm0\bm1'} & = e^{-2 \mbox{i}\omega}\bme_{\bm1 \bm0'}.
\end{align*}
The Lorentz transformation and conformal factor $\kappa$ relating the
frames is given by
\begin{equation}
\label{LorentzTransformationExplicitScriMinus}
\Lambda^{\bm0}{}_{\bm1}=\frac{2 e^{\mbox{i}\omega}}{\sqrt{\rho}(1-\tau)}\qquad
\Lambda^{\bm1}{}_{\bm0}= \frac{
  e^{-\mbox{i}\omega}\sqrt{\rho}(1-\tau)}{2}, \qquad
\Lambda^{\bm0}{}_{\bm0}=\Lambda^{\bm1}{}_{\bm1}=0, \quad \qquad
\kappa=1.
\end{equation}
\end{proposition}

\begin{proposition}\label{PropositionNPconstantsGeneralStructureTimeDual}
The electromagnetic constants $\bar{F}_{m}^{n}{}$ at
$\mathscr{I}^{-}$ are given by
\[
\bar{F}_{m}^{n-1}=Q^{-}(n)D_{n+1,n,m},
\]
where $Q^{-}(n)$ is a numerical coefficient.
\end{proposition}

Finally, recalling the results of Propositions
 \ref{PropositionNPconstantsGeneralStructure}, and
 \ref{PropositionNPconstantsGeneralStructureTimeDual},  the
 expressions for $C_{p,\ell,m}$ and $D_{p,\ell,m}$ given in Lemma
 \ref{MaxwellSolutionsExplicit} and Remark
 \ref{TimeSymmetricInitialData} one obtains the following:

\begin{theorem}\label{TheoremEMconstants}
  Initial data for the spin-1 field (Maxwell's equations)
  on the Minkowski spacetime,
  satisfying the regularity condition of Assumption
  \ref{RegularityConditionMaxwell} give rise to a solution $\phi_{AB}$
  whose electromagnetic NP constants $F_{m}^{n}{}$ and
  $\bar{F}_{m}^{n}$ at $\mathscr{I}^{+}$ and $\mathscr{I}^{-}$, that
  in general, do not coincide. If the data is time-symmetric, the
  electromagnetic NP constants at $\mathscr{I}^{+}$ and
  $\mathscr{I}^{-}$, correspond to the same part of the initial data
  and, hence, up to a numerical factor $Q^{+}(n)/Q^{-}(n)$, coincide.
\end{theorem}

\section{The NP constants for the massless spin-2 field}
\label{LinearisedGravityNPConstants}

In this section an analogous analysis to that given in Section
 \ref{TheElectromagneticNPConstants} 
is performed for the case of the spin-2 massless field. The same notation 
as the one introduced in Section  \ref{TheElectromagneticNPConstants}
 will be used. In particular, the spin dyads $\tilde{\epsilon}_{\bmA}{}^{A}$, $\epsilon^{'}_{\bmA}{}^{A}$ and $\epsilon_{\bmA}{}^{A}$ associated to 
$\tilde{\bm\eta}$, $\bmg'$ and $\bmg$ will be employed.
 To introduce the gravitational NP constants originally
 introduced in \cite{NewPen68},  let 
$\tilde{\phi}_{0}$, $\tilde{\phi}_{1}$, $\tilde{\phi}_{2}$, $\tilde{\phi}_{3}$
and $\tilde{\phi}_{4}$ denote the components of the spin-2 massless field
 $\tilde{\phi}_{ABCD}$ respect to $\tilde{\epsilon}_{\bmA}{}^{A}$. The spin-2 equation reads
\begin{equation}\label{BianchiPhysical}
\tilde{\nabla}_{A'}{}^{A}\tilde{\phi}_{ABCD}=0.
\end{equation}

\begin{assumption}
{\em Following  \cite{NewPen68}, the component ${\phi}_{0}$ is assumed to have the 
expansion
\begin{equation}
\label{PeelingPhysicalWeyl}
\tilde{\phi}_{0}=\sum_{n=0}^{N}\frac{\tilde{\phi}^{n}_{0}}{\tilde{\rho}^{5+n}}
+ \mathcal{O}\Big(\frac{1}{\tilde{\rho}^{5+N}}\Big),
\end{equation}
where the coefficients $\tilde{\phi}_{0}^{n}$ do not depend on 
$\tilde{\rho}$. }
\end{assumption}

As already mentioned, the field $\tilde{\phi}_{ABCD}$ provides
a description of the linearised gravitational field over the Minkowski
spacetime. 
In the full non-linear theory, the linear field 
$\tilde{\phi}_{ABCD}$ is replaced by the Weyl spinor $\Psi_{ABCD}$ and the
analogue of equation \eqref{BianchiPhysical}
encodes the second Bianchi identity in vacuum ---see \cite{NewPen68}.
The spin-2 NP quantities are defined through the following integrals
over cuts $\mathcal{C}$ of null infinity:
\begin{equation*}
G_{m}^{n}\equiv \int_{\mathcal{C}} \bar{Y}_{2;n+2,m}\tilde{\phi}_{0}^{n+1}\mbox{d}S,
\end{equation*}
where $n, m \in \mathbb{Z}$ with $n \geq 0$,
$|m|\leq n+2$ and $\mbox{d}S$ denotes the area element 
respect to $\bm\sigma$. The NP constants $G_{m}^{n}$
 are absolutely conserved in the sense that their 
value is independent on the cut
$\mathcal{C}$ on which they are evaluated.
\begin{remark}
{\em In particular, the constants $G^{0}_m$ are also conserved in
  the full non-linear case of the gravitational field 
where  $\tilde{\phi}_0$ is replaced by the component $\Psi_0$ of the Weyl
spinor ${\Psi}_{ABCD}$ 
 ---see \cite{NewPen68}. These are the only constants of the hierarchy
which are generically inherited in the non-linear case.}
\end{remark}

\subsection{Translation to the F-gauge}

An expression for the gravitational NP constants in the
F-gauge has been given in Section III of
\cite{FriKan00}. In order to provide a self-contained discussion and
for the ease of comparison with the analysis made in 
Section  \ref{TheElectromagneticNPConstants} the analogue of 
 Formula (III.5) of \cite{FriKan00} will
be derived in accordance with the notation and conventions used in this article.
In view of equation \eqref{gprimeTophysicalMinkowski}, one has that,
as a consequence of the standard conformal transformation law
for the spin-2 equation ---see \cite{Ste91},
   the spinor $\phi'_{ABCD}$, satisfying
\[
\nabla'_{A'}{}^{A}\phi'_{ABCD}=0,
\] where $\nabla'_{AA'}$ is the Levi-Civita connection of $\bmg'$,
 is related to  ${\phi}_{ABCD}$ via
\begin{equation}
\label{UnphysicalWeylAndPhysicalWeylSpinor}
\phi'_{ABCD}=\tilde{\rho}\tilde{\phi}_{ABCD}.
\end{equation}
Therefore, using equations
 \eqref{PeelingPhysicalWeyl},\eqref{RelationBetweenSpinDyads}
and  \eqref{UnphysicalWeylAndPhysicalWeylSpinor}, 
 one obtains
\begin{equation*}
 \phi'_{0}=\sum_{n=0}^{N}\frac{\tilde{\phi}^{n}_{0}}{\tilde{\rho}^n} 
+ \mathcal{O}\Big(\frac{1}{\tilde{\rho}^{N}}\Big),
\end{equation*}
where $\phi'_{0}\equiv\phi'_{ABCD}o'^{A}o'^{B}o'^{C}o'^{D}$. Using the last expansion and recalling that
$\bme'_{\bm0\bm0'}=-\tilde{\rho}^2\bm\partial_{\tilde{\rho}}$
one obtains, after consecutive applications of $\bme'_{\bm0 \bm0'}$,
the expression 
\begin{equation}
G^{n}_{m}=\frac{1}{(n+1)!}\int_{\mathcal{C}}\bar{Y}_{2;n+2,m}\bme'^{(n+1)}_{\bm0\bm0'}(\phi'_{0})\mbox{d}S.\label{GravNPquantitiesInNPgauge}
\end{equation}

To derive an expression for the spin-2 NP
constants in the F-gauge one recalls the relation
between the  $\bmg'$ and $\bmg$ representations and their
associated spin dyads encoded  in equation \eqref{RelationSpinDyadsEqs}.
Once again, as a consequence of the conformal transformation laws
for the spin-2 equation one has that the spinor field
$\phi_{ABCD}$ related to $\phi'_{ABCD}$ through
\[
\phi'_{ABCD}=\kappa^{-1}\phi_{ABCD},
\]
satisfies
\[
\nabla_{A'}{}^{A}\phi_{ABCD}=0,
\]
where $\nabla_{AA'}$ represents the Levi-Civita connection respect to
$\bmg$. Additionally, one has that
\[
\phi'_{0}=\kappa^{-3}\Lambda^{\bmA}{}_{\bm0}\Lambda^{\bmB}{}_{\bm0}
\Lambda^{\bmC}{}_{\bm0}\Lambda^{\bmD}{}_{\bm0}\phi_{\bmA \bmB \bmC \bmD},
\]
where $\phi_{\bmA \bmB \bmC \bmD}\equiv\epsilon_{\bmA}{}^{A}\epsilon_{\bmB}{}^{B}
\epsilon_{\bmC}{}^{C}\epsilon_{\bmD}{}^{D}\phi_{ABCD}$ .

\subsubsection{Explicit computation of the first constant}
Using equation \eqref{GravNPquantitiesInNPgauge} and the Leibniz rule one 
obtains the analogue of Equation (III.5) of \cite{FriKan00} written in accordance with
the notation and conventions used in this article
\begin{multline}
  G^{0}_{m}=\int_{\mathcal{C}}\bar{Y}_{2;2,m}\kappa^{-4} 
       \Big( \Lambda^{\bmA}{}_{\bm0}\Lambda^{\bmB}{}_{\bm0}
\Lambda^{\bmC}{}_{\bm0}\Lambda^{\bmD}{}_{\bm0}\big(\Lambda^{\bmE}{}_{\bm0} \bar{\Lambda}^{\bmE'}{}_{\bm0'}\bme_{\bmE \bmE'}(\phi_{\bmA \bmB \bmC \bmD})  \\
 - 3 \phi_{\bmA \bmB \bmC\bmD}\bme'_{\bm0\bm0'}(\kappa)\big) 
 + 4 \kappa \Lambda^{\bmA}{}_{\bm0}\Lambda^{\bmB}{}_{\bm0}
\Lambda^{\bmC}{}_{\bm0}\phi_{\bmA \bmB \bmC\bmD}\bme'_{\bm0 \bm0'}(\Lambda^{\bmD}{}_{\bm0})\Big) \mbox{d}S.\label{EquationIII5}
\end{multline}
 
Particularising the discussion to the case of the Minkowski spacetime,
simplifies the expressions considerably.  To see this, observe that,
using Proposition \ref{PropNPtoFgauge} and the notation introduced in
equation \eqref{SimplifyingNotation}, a direct calculation reveals that
$\phi_0'=\Lambda^4\phi_4$.  Using this relation, and the short-hands
introduced in equation \eqref{SimplifyingNotation} one has that
\begin{equation}
\label{FirstDerivativeExplicit}
\bme'_{\bm0\bm0'}(\phi_{0}')=\frac{\sqrt{2}}{2}\Lambda^6(\bme \phi_4 -2\phi_4).
\end{equation}
Using that
\begin{equation}\label{bmea4rhop}
  \bme(a_{4,p,\ell,m}(\tau)\rho^p)= ((1+\tau)\dot{a}_{4,p,l,m}(\tau)-p
  a_{4,p,\ell,m}(\tau))\rho^p,
\end{equation}
and the Ansatz \eqref{ExpansionGravPhin} a calculation gives
\begin{equation*}
  \bme'_{\bm0\bm0'}(\phi_{0}') = \frac{\sqrt{2}}{2}
  \sum_{p=2}^{\infty}\sum_{\ell=2}^{p}\sum_{m=-\ell}^{\ell}
  \frac{1}{p!}\Lambda^6 \rho^p ((1+\tau)\dot{a}_{4,p,l,m}(\tau)-(p+2)
  a_{4,p,\ell,m}(\tau))Y_{2;\ell, m}.
\end{equation*}
Separating the dependence in $\tau$ and $\rho$ explicitly, the latter
expression can be written as
\begin{equation}\label{FirstDerivativeSpin2KeyExpression}
  \bme'_{\bm0\bm0'}(\phi_{0}') = 2^6 \Big(\frac{\sqrt{2}}{2}\Big)
  \sum_{p=2}^{\infty}\sum_{\ell=2}^{p}\sum_{m=-\ell}^{\ell}
  \frac{1}{p!}(1+\tau)^{-6}
  \rho^{(p-3)}A^{1}_{4,p,l,m}(\tau)Y_{2;\ell, m},
\end{equation}
where
\begin{align}\label{FirstOperatorSpin2}
A^{1}_{4,p,l,m}(\tau) \equiv ((1+\tau)\dot{a}_{4,p,l,m}(\tau)-(2+p)
a_{4,p,\ell,m}(\tau)).
\end{align}
 Using equations
  \eqref{FirstDerivativeSpin2KeyExpression} and
  \eqref{GravNPquantitiesInNPgauge} computing the first set of NP
  constants reduces to the evaluating the following
  integral
  \begin{align}
  \label{FirstConstantsSpin2SimplifiedStepThree}
G_{m}^{0}=2^6 \Big(\frac{\sqrt{2}}{2}\Big)\lim_{\substack{\rho \to \rho_\star \\ \tau \to 1}}\Bigg(
\int_{\mathbb{S}^2}
\sum_{p=2}^{\infty}\sum_{\ell=2}^{p}\sum_{m=-\ell}^{\ell}
\frac{1}{p!}(1+\tau)^{-6} \rho^{(p-3)}A^{1}_{4,p,l,m}(\tau)Y_{2;\ell,
  m}\bar{Y}_{2;2,m} \mbox{d}S \Bigg).
  \end{align}
 Using the orthogonality relation \eqref{OrthogonalitySphericalHarmonics}
one obtains
\begin{align}\label{PreFirstNPSpin2Constant}
G_{m}^{0}=2^6 \Big(\frac{\sqrt{2}}{2}\Big) \sum_{p=2}^{\infty}
\frac{1}{p!}\rho_{\star}^{(p-3)}A^{1}_{4,p,2,m}|_{\mathscr{I}^{+}}.
\end{align}
A direct calculation using the explicit form of $a_{4,p,l,m}(\tau)$
gives the following:
\begin{remark}\label{remarkMagicCancelationFirstDerivativeSpin2}
   $A^{1}_{4,p,1,m}|_{\mathscr{I}^{+}} = 0$ for $p=2$ and $p \geq 4$.
\end{remark}
Exploiting Remark \ref{remarkMagicCancelationFirstDerivativeSpin2} a
calculation one concludes that,
\begin{align}\label{FirstNPSpin2Constant}
G_{m}^{0}= -\frac{80\sqrt{2}}{3} C_{3,2,m}.
\end{align}
Completely analogous remarks as those made for the calculation of the
electromagnetic NP constants can be made in this case: the constant
comes from the $p=3$ term and all the terms with $p\geq 4$ that could
potentially contribute to the NP constant identically vanish by virtue
of Remark \ref{remarkMagicCancelationFirstDerivativeSpin2}. The
regularity condition of Assumption \ref{RegularityConditionSpin2} has
been used to have well defined classical NP constants. As in the
electromagnetic case, one can avoid using the results of Remark
\ref{remarkMagicCancelationFirstDerivativeSpin2} by simply invoking
the finiteness and constancy of the NP constants and evaluating at the
cut $\mathcal{C}=I^{+}$.

\subsubsection{The general case}
 Revisiting the calculation of $G_{m}^{0}$ one can see that the mechanism is completely analogous to the electromagnetic case:

\begin{lemma}\label{LemmaDerivativesExplicitSpin2GeneralStructure}
For any integer $n \geq 1$ 
\begin{equation}\label{nDerivativeSpin2KeyExpression}
  \bme'{}_{\bm0\bm0'}^{(n)}(\phi_{0}') = 2^{2(n+2)} \bigg(\frac{\sqrt{2}}{2}\bigg)^n
  \sum_{p=2}^{\infty}\sum_{\ell=2}^{p}\sum_{m=-\ell}^{\ell}
  \frac{1}{p!}(1+\tau)^{-2(n+2)}
  \rho^{p-(n+2)}A^{n}_{4,p,l,m}(\tau)Y_{2;\ell, m},
\end{equation}
with,
\begin{equation}\label{nOperatorSpin2}
  A^{n}_{4,p,l,m}(\tau) \equiv \sum_{k=0}^{n}(-1)^k {n \choose k}\frac{(k+p+1)!}{(p+1)!}(1+\tau)^{n-k}a^{(n-k)}_{4,p,l,m},
\end{equation}
where
$a^{(k)}_{4,p,l,m} \equiv (\bm\partial_\tau)^{k}a_{4,p,l,m}$.
\end{lemma}

\begin{proof}
  To prove this result one proceeds by induction.  Equations
  \eqref{FirstDerivativeSpin2KeyExpression} and
  \eqref{FirstOperatorSpin2} already show that the result is valid for
  $n=1$.  This constitutes the basis of induction.  Now, assume that equations \eqref{nDerivativeSpin2KeyExpression} and
  \eqref{nOperatorSpin2} hold (induction hypothesis), then, applying
  $\bme'_{\bm0\bm0'}$ to equation
  \eqref{nDerivativeSpin2KeyExpression} renders

\begin{multline}\label{nplusoneDerivativeSpin2KeyExpression}
  \bme'{}_{\bm0\bm0'}^{(n+3)}(\phi_{0}') = 2^{2(n+3)}
  \bigg(\frac{\sqrt{2}}{2}\bigg)^{(n+3)}
  \sum_{p=2}^{\infty}\sum_{\ell=2}^{p}\sum_{m=-\ell}^{\ell}
  \frac{1}{p!}(1+\tau)^{-2(n+3)}
  \rho^{p-(n+3)}R^{n}_{4,p,l,m}(\tau)Y_{2;\ell, m}  ,
\end{multline}
where
\begin{equation*}
R^{n}_{4,p,l,m}(\tau) = (1+\tau)\dot{A}^{n}_{4,p,l,m}(\tau) - (p+n+2)
A^{n}_{4,p,l,m}(\tau).
\end{equation*}
Substituting \eqref{nOperatorSpin2} into the last expression and
omitting the subindices gives
\begin{multline*}
  R^{n}(\tau)= (1+\tau)\sum_{k=0}^{n}(-1)^k {n \choose
    k}\frac{(k+p+1)!}{(p+1)!}\Big((n-k)(1+\tau)^{n-k-1}a^{(n-k)} +
  (1+\tau)^{n-k}a^{(n-k-1)}\Big) \\ -(p+n+2)\sum_{k=0}^{n}(-1)^k {n
    \choose k}\frac{(k+p+1)!}{(p+1)!}(1+\tau)^{n-k}a^{(n-k)}.
\end{multline*}
Reorganising the last expression one gets
\begin{multline}\label{RnIntermediateSpin2}
  R^{n}(\tau)= \sum_{k=0}^{n}(-1)^k {n \choose
    k}\frac{(k+p+1)!}{(p+1)!}(1+\tau)^{n-k+1}a^{(n-k+1)}
  \\ -\sum_{k=0}^{n}(-1)^k {n \choose k}\frac{(k+p+1)!}{(p+1)!}(p+k+2)(1+\tau)^{n-k}a^{(n-k)}.
\end{multline}
Separating the first term in the first sum of
equation \eqref{RnIntermediateSpin2} gives
\begin{multline*}
  R^{n}(\tau)= (1+\tau)^{n+1}a^{(n+1)} +
  \sum_{\lambda=0}^{n-1}(-1)^{\lambda +1} {n \choose \lambda + 1
  }\frac{(p+\lambda +2)!}{(p+1)!}(1+\tau)^{n-\lambda}a^{(n-\lambda)}
  \\ +\sum_{k=0}^{n}(-1)^{k+1} {n \choose
    k}\frac{(p+k+2)!}{(p+1)!}(1+\tau)^{n-k}a^{(n-k)}.
\end{multline*}
Expanding the last term in the second sum and rearranging renders
\begin{multline*}
  R^{n}(\tau)= (1+\tau)^{n+1}a^{(n+1)} +
  \sum_{k=0}^{n-1}(-1)^{k+1}\frac{(p+k
    +2)!}{(p+1)!}(1+\tau)^{n-k}a^{(n-k)}\Bigg( {n \choose k }
  + {n \choose  n+1 } \Bigg) \\ + (-1)^{n+1}\frac{(p+n+2)!}{(p+1)!} a.
\end{multline*}
Applying the recursive identity of the binomial coefficients
\eqref{BinomialRecursive} one gets
\begin{multline*}
  R^{n}(\tau)= (1+\tau)^{n+1}a^{(n+1)} +
  \sum_{k=0}^{n-1}(-1)^{k+1}{n+1 \choose k+1 }\frac{(p+k
    +2)!}{(p+1)!}(1+\tau)^{n-k}a^{(n-k)}\\
  + (-1)^{n+1}\frac{(p+n+2)!}{(p+1)!} a.
\end{multline*}
The last expression can be rewritten as
\begin{multline*}
  R^{n}(\tau)= (1+\tau)^{n+1}a^{(n+1)} +
  \sum_{i=0}^{n-1}(-1)^{i}{n+1 \choose i
  }\frac{(p+i+1)!}{(p+1)!}(1+\tau)^{n-i-1}a^{(n-i-1)}\\ +
  (-1)^{n+1}\frac{(p+n+2)!}{(p+1)!} a.
\end{multline*}
Reshuffling the counter to absorb the first and last term and using
the definition given in equation \eqref{nOperatorSpin2} renders
\begin{equation*}
R^{n}_{4,p,\ell,m}(\tau)=A^{n+1}_{4,p,\ell,m}(\tau).
\end{equation*}
Finally, substituting the last expression into
equation \eqref{nplusoneDerivativeSpin2KeyExpression} one concludes that
\begin{multline*}
  \bme'{}_{\bm0\bm0'}^{(n+1)}(\phi_{0}') = 2^{2(n+3)}
  \bigg(\frac{\sqrt{2}}{2}\bigg)^{n+1}
  \sum_{p=2}^{\infty}\sum_{\ell=2}^{p}\sum_{m=-\ell}^{\ell}
  \frac{1}{p!}(1+\tau)^{-2(n+3)}
  \rho^{p-(n+3)}A^{n+1}_{4,p,l,m}(\tau)Y_{2;\ell, m}  ,
\end{multline*}
which completes the induction step.
\end{proof}
Exploiting Lemma \ref{LemmaDerivativesExplicitSpin2GeneralStructure}
 to determine the general structure of the NP constants is
straightforward:
\begin{proposition}\label{PropositionSpin2NPconstantsGeneralStructure}
  The  NP constants in Minkowski spacetime $G^{n-1}_{m}$ are given by
  \begin{equation*}
    G_{m}^{n-1}= Q^+(n) C_{n+2,n+1,m}.
\end{equation*}
  where $Q^+(n)$ is a numerical factor and
  $C_{n+1,n,m}$ is determined by the initial data $a_{0,n+1,n,m}(0)$
  and $a_{4,n+1,n,m}(0)$ as given in Lemma \ref{Spin2SolutionsExplicit}.
\end{proposition}
\begin{proof}
The orthogonality condition
\eqref{OrthogonalitySphericalHarmonics}, along with
the expression for the NP constants given in equation
\eqref{GravNPquantitiesInNPgauge} and the equations 
\eqref{nDerivativeSpin2KeyExpression} and \eqref{nOperatorSpin2} render
\begin{align}\label{PrenthNPSpin2Constant}
 G_{m}^{n-1}= 2^{n+2} \bigg(\frac{\sqrt{2}}{2}\bigg)^n
  \Big(\frac{1}{n!}\Big) \sum_{p=2}^{\infty}
\frac{1}{p!}\rho_{\star}^{(p-(n+2))}A^{n}_{4,p,n+1,m}|_{\mathscr{I}^{+}}.
\end{align}
Invoking the finiteness and constancy of the NP constants in order to
evaluate the last expression on the cut $I^+$ one concludes that
\begin{align}\label{nthNPMaxwellConstant}
G_{m}^{n-1}= 2^{n+2}
\bigg(\frac{\sqrt{2}}{2}\bigg)^n\bigg(\frac{1}{n!(n+2)!}\bigg)A^{n}_{4,n+2,n+1,m}|_{\mathscr{I}^{+}}.
\end{align}
direct inspection of $A^{n}_{4,p,n+2,m}|_{\mathscr{I}^{+}}$ renders 
\begin{align}\label{nNPSpin2ConstantInTermsofInitialData}
G_{m}^{n-1}= Q^+(n) C_{n+2,n+1,m},
\end{align}
where $Q^+(n)$ is an irrelevant numerical factor.
\end{proof}

\subsection{The constants at $\mathscr{I}^-$}
The time dual result can be obtained in a similar
way as it was done for the in the electromagnetic case.
Using Propositions \ref{PropNPtoFgaugePast} and
\ref{PropositionSpin2NPconstantsGeneralStructure}, one obtains:

\begin{proposition}\label{PropositionSpin2NPconstantsGeneralStructureTimeDual}
The NP constants $\bar{G}_{m}^{n}{}$ at
$\mathscr{I}^{-}$ are given by
\[
\bar{G}_{m}^{n-1}=Q^{-}(n)D_{n+2,n+1,m},
\]
where $Q^{-}(n)$ is a numerical coefficient.
\end{proposition}

Recalling the results of Propositions
 \ref{PropositionSpin2NPconstantsGeneralStructure},
 \ref{PropositionSpin2NPconstantsGeneralStructureTimeDual},  the
 expressions for $C_{p,\ell,m}$ and $D_{p,\ell,m}$ given in Lemma
 \ref{Spin2SolutionsExplicit} and Remark
 \ref{TimeSymmetricInitialDataSpin2} one obtains the following: 

\begin{theorem}\label{TheoremSpin2constants}
   Initial data for the spin-2 field on the Minkowski spacetime,
   satisfying the regularity condition of Assumption
   \ref{RegularityConditionSpin2} give rise to a solution
   $\phi_{ABCD}$ whose associated NP constants $G_{m}^{n}{}$ and
   $\bar{G}_{m}^{n}$ at $\mathscr{I}^{+}$ and $\mathscr{I}^{-}$, that in
   general, do not coincide. If the data is time-symmetric, the NP
   constants at $\mathscr{I}^{+}$ and $\mathscr{I}^{-}$, correspond to
   the same part of the initial data and, hence, up to a numerical
   factor $Q^{+}(n)/Q^{-}(n)$, coincide.
\end{theorem}

\begin{remark}
{\em A similar symmetric behaviour has been observed in the
  gravitational case in \cite{Val07a}. In that reference the
  Newman-Penrose constants at future and past null infinity of the
  spacetime arising from Bowen-York initial data have been computed.}
\end{remark}

\section{Conclusions}
\label{Conclusions}
In this article the correspondence between initial data given on a
Cauchy hypersurface $\mathcal{S}$ intersecting $i^{0}$ on Minkowski
spacetime for the spin-1 (electromagnetic) and spin-2 fields and their
associated NP constants is analysed.  This analysis has been done for
the full hierarchy of NP constants $F^{n}_{m}$ and $G_{m}^{n}$ in the
Minkowski spacetime.

\medskip

For the electromagnetic case, it was shown that, once the initial data
for the Maxwell spinor is written as an expansion of the form
\eqref{InitialExpansionPhin}, the electromagnetic NP constants
$F^{n-1}_{m}{}$ at $\mathscr{I}^{+}$ can be identified with a
particular combination ---denoted as $C_{p,\ell,m}$--- of the initial
data $a_{0,p;\ell,m}(0)$ and $a_{2,p;\ell,m}(0)$ with $p=n+1$ and
$\ell=n$ ---see Lemma \ref{MaxwellSolutionsExplicit} and Proposition
\ref{PropositionNPconstantsGeneralStructure}.  Since $1 \leq \ell \leq
p$, one concludes that $F^{n}_{m}$ are in correspondence with the
second highest harmonic but are insensitive to the initial data for
 lower modes $\ell \leq p-2$ or the highest mode $p=\ell$. Notice,
that it has been assumed that the logarithmic terms appearing at
$p=\ell$ are not present in the solution which, of course, restricts the
initial data ---see Assumption \ref{RegularityConditionMaxwell}.
This restriction is necessary, otherwise the NP constants are not
well defined ---see discussion in \cite{Val98} and
\cite{Val99a}.
In an analogous way, one can identify the electromagnetic NP constants
$\bar{F}^{n-1}_{m}$ at $\mathscr{I}^{-}$ with a different combination
---denoted as $D_{p,\ell,m}$--- of the initial data $a_{0,p;\ell,m}(0)$
and $a_{2,p;\ell,n}(0)$ and similar remarks apply. The crucial
observation to be made is that in general $C_{p,\ell\,m} \neq
D_{p,\ell\,m}$, and hence, the NP constants at $\mathscr{I}^{+}$  and
$\mathscr{I}^{-}$ do not coincide ---see Theorem
\ref{TheoremEMconstants}. Notice however that if the initial data
considered is time-symmetric then, $C_{p,\ell,m}=-D_{p,\ell,m}$, and
consequently, there is a correspondence between the NP constants defined at
$\mathscr{I}^{+}$ and $\mathscr{I}^{-}$.

\medskip
The same analysis was performed for a spin-2 field on a Minkowski
background. The conclusion being that the NP constants $G^{n-1}\_{m}$
at $\mathscr{I}^+$ and $\mathscr{I}^-$ depend on the initial data for
$a_{0,p;\ell,m}(0)$ and $a_{4,p;\ell,m}(0)$ with $p=n+2$ and
$\ell=n+1$ via the expressions for $C_{p,\ell,m}$ and $D_{p,\ell,m}$
as given in Lemma \ref{Spin2SolutionsExplicit} ---see Propositions
\ref{PropositionSpin2NPconstantsGeneralStructure} and
\ref{PropositionSpin2NPconstantsGeneralStructureTimeDual}.  The main
point to be stressed about the particular form of $C_{p,\ell,m}$ and
$D_{p,\ell,m}$ is that $C_{p,\ell,m} \neq D_{p,\ell,n}$ ---see Theorem
\ref{TheoremSpin2constants}.  As in the electromagnetic case, in order
to have well defined NP constants it is necessary to assume that the
initial data satisfy the regularity condition of Assumption
\ref{RegularityConditionSpin2} ---see \cite{Val03a}, \cite{Fri98a} and
\cite{CFEBook} for further discussion on this regularity condition. The overall conclusion is then then that, for \emph{truly generic initial data}
the \emph{classical NP constants}
are not defined and even if one considers data satisfying the
regularity condition of Assumption \ref{RegularityConditionSpin2}
---or Assumption \ref{RegularityConditionMaxwell}
for the electromagnetic case--- , the NP
constants at $\mathscr{I}^{+}$ and $\mathscr{I}^{-}$ do not
coincide. However, restricting the initial data to be time-symmetric
is a sufficient condition to achieve this
correspondence.

\subsection*{Acknowledgements}

We have profited from discussions with D. Hilditch, M. Magdy Ali Mohamed,
M. Minnuci and G. Taujanskas.  EG
gratefully acknowledges the support from Consejo Nacional de Ciencia y
Tecnolog\'ia (CONACyT Scholarship 494039/218141) in the early stages
of this article and The European Union’s H2020 ERC Consolidator Grant
“Matter and Strong-Field Gravity: New Frontiers in Einstein’s Theory,”
Grant Agreement No. MaGRaTh-646597, the PO FEDER-FSE Bourgogne 2014/2020 program as well as
 the EIPHI Graduate School (contract ANR-17-EURE-0002) as part of the ISA 2019 project, during its completion.
EG and JAVK, thank, respectively, the
hospitality of the Gravitational Physics Group of the Faculty of
Physics of the University of Vienna and the GRIT group at CENTRA of
Instituto Superior T\'ecnico during a scientific visit.

\appendix
\section{The connection on $\mathbb{S}^2$}
\label{ConnectionTwoSphere}

In this section expressions for the connection coefficients ---of the
Levi-Civita connection--- respect to a complex null frame which do not
make reference to any particular coordinate system on $\mathbb{S}^2$
are obtained. To set up the notation, it is convenient to start the
discussion  writing the  Cartan structure equations
 in accordance with the conventions used in this article:
\begin{subequations}
\begin{align}
&  \mathbf{d}\bm\omega^{\bma}=-\bm\gamma^{\bma}{}_{\bmb}\wedge\bm\omega^{\bmb} ,
\label{FirstCartanStructureEquation}  \\ &
  \mathbf{d}\bm\gamma^{\bma}{}_{\bmb}=-\bm\gamma^{\bma}{}_{\bmd}\wedge\bm\gamma^{\bmd}{}_{\bmb}
  + \bm\Omega^{\bma}{}_{\bmb}. \label{SecondCartanStructureEquation}
\end{align}
\end{subequations}
In the last expressions the curvature 2-form $\bm\Omega^{\bma}{}_{\bmb}$
 and connection 1-form $\bm\gamma^{\bmc}{}_{\bmb}$ are defined via
\begin{equation}
\label{DefCurvatureAndConnectionForm}
\bm\Omega^{\bma}{}_{\bmb} \equiv
\frac{1}{2}R^{\bma}{}_{\bmb\bmc\bmd}\bm\omega^{\bmc}\wedge\bm\omega^{\bmd},
 \qquad \qquad
\bm\gamma^{\bmc}{}_{\bmb} \equiv \Gamma_{\bma}{}^{\bmc}{}_{\bmb}\wedge
 \bm\omega^{\bma}.
\end{equation}
The connection coefficients $\Gamma_{\bma}{}^{\bmc}{}_{\bmb}$ of the Levi-Civita
connection $\nabla$ respect to a given frame $\bme_{\bma}$ are
defined as $\Gamma_{\bma}{}^{\bmc}{}_{\bmb} \equiv \langle
\bm\omega^{\bmc}, \nabla_{\bma}\bme_{\bmb}\rangle$.

\medskip
In the remaining part of this appendix the discussion
is particularised to the case of $\mathbb{S}^2$.
  Let $\{\bm\partial_{+}, \bm\partial_{-} \}$ be a
complex null frame on $\mathbb{S}^2$ with corresponding dual covectors
$\{\bm\omega^{+},\bm\omega^{-} \}$.  Namely, one considers
\[ \bm\sigma=2(\bm\omega^{+}\otimes
 \bm\omega^{-}+\bm\omega^{-}\otimes \bm\omega^{+}),
 \qquad \bm\sigma^{\flat}= \frac{1}{2}(\bm\partial_{+}\otimes
 \bm\partial_{-} + \bm\partial_{-}\otimes \bm\partial_{+}), \] where
 $\bm\sigma$ and $\bm\sigma^{\flat}$ denote the covariant and
 contravariant version of the standard metric on
 $\mathbb{S}^2$. Furthermore, one assumes that
\begin{equation} \label{ConjugationBasis}
\bm\partial_{+}= \overline{\bm\partial_{-}},
\end{equation}
 and consequently $\bm\omega^{+}=\overline{\bm\omega^{-}}$.  To start
 the discussion observe that $[\bm\partial_{+},\bm\partial_{-}]$ and
 its complex conjugate can be expressed as a linear combination of the
 basis vectors $\bm\partial_{+}$ and $\bm\partial_{-}$.  A direct
 inspection, taking into account the condition encoded in equation
 \eqref{ConjugationBasis}, reveals that
\begin{equation}\label{Commutator}
[\bm\partial_{+},\bm\partial_{-}]=\varpi
\bm\partial_{+}-\overline{\varpi}\bm\partial_{-},
\end{equation}
where $\omega$ is a scalar field over $\mathbb{S}^2$.  Using the 
no-torsion condition of the Levi-Civita connection $\slashed{\nabla}$ on
$\mathbb{S}^2$ we get from equation \eqref{Commutator} that
\begin{equation}\label{NotorsionTwoSphere}
\slashed{\nabla}_{+}\bm\partial_{-}-\slashed{\nabla}_{-}\bm\partial_{+}=\varpi
\bm\partial_{+}-\overline{\varpi}\bm\partial_{-},
\end{equation}
where $\slashed{\nabla}_{+}$ and $\slashed{\nabla}_{-}$ denote a
covariant derivative in the direction of $\bm\partial_{+}$ and
$\bm\partial_{-}$ respectively.  Using equation \eqref{NotorsionTwoSphere}
 and the metricity conditions $\slashed{\nabla}_{+}\bm\sigma =0,
\slashed{\nabla}_{-}\bm\sigma=0$, one finds that the only non-zero
connection coefficients are all encoded in the scalar field $\omega$:
\[
\Gamma_{-}{}^{-}{}_{-}=\overline{\Gamma_{+}{}^{+}{}_{+}}
=-\Gamma_{-}{}^{+}{}_{+}=-\overline{\Gamma_{+}{}^{-}{}_{-}}=\varpi.
\]
The connection can be compactly encoded in the curvature 1-form
$\bm\gamma^{\bma}{}_{\bmb}$ as defined in equation
\eqref{DefCurvatureAndConnectionForm}.
A direct computation renders
\begin{equation*}
\bm\gamma^{+}{}_{+}=\overline{\bm\gamma^{-}{}_{-}} =
\overline{\varpi}\bm\omega^{+}-\varpi\bm\omega^{-}, \qquad
\bm\gamma^{+}{}_{-}=\bm\gamma^{-}{}_{+}=0.
\end{equation*}
Using the first Cartan structure equation encoded in 
 \eqref{FirstCartanStructureEquation}, one obtains
\begin{equation}\label{ExtDerOmegaPlusMinus}
\mathbf{d}\bm\omega^{+}=-\varpi\bm\omega^{+}\wedge\bm\omega^{-},
\qquad
\mathbf{d}\bm\omega^{-}=\overline{\varpi}\bm\omega^{+}\wedge\bm\omega^{-}.
\end{equation}
For completeness, using the above expressions and the second Cartan
structure equation encoded in \eqref{SecondCartanStructureEquation},
 one can directly compute the
curvature form $\bm\Omega^{\bma}{}_{\bmb}$:
\[
\bm\Omega^{+}{}_{+} =\overline{\bm\Omega^{-}{}_{-}}=-2(|\varpi|^2 +
\frac{1}{2}(\bm\partial_{+}\varpi +
\bm\partial_{-}\overline{\varpi}))\bm\omega^{+}\wedge\bm\omega^{-}.
\]
Notice that, in order to find further information about $\omega$ one
can exploit the fact that the Riemann curvature for maximally
symmetric spaces $(\mathcal{N},\bmh)$ is given by
\[
R_{\bma\bmb\bmc\bmd}=\frac{1}{2}R(h_{\bma\bmb}h_{\bmb\bmd}-h_{\bma\bmd}h_{\bmb\bmc}),
\]
where $R$ is the Ricci scalar of the Levi-Civita connection of the
metric $\bmh$ on $\mathcal{N}$. Since the Ricci scalar for
$\mathbb{S}^2$ is $R=-2$, using equation \eqref{DefCurvatureAndConnectionForm}
 one
finds that
\[
\bm\Omega^{+}{}_{+}=\overline{\bm\Omega^{-}{}_{-}}=2
\bm\omega^{+}\wedge\bm\omega^{-}.
\]
Consequently, one concludes that the scalar field $\omega$ satisfies
\begin{equation}
\label{OmegaIsCotangent}
|\varpi|^2 + \frac{1}{2}(\bm\partial_{+}\varpi +
\bm\partial_{-}\overline{\varpi})=-1.
\end{equation}

\section{The $\eth$ and $\bar{\eth}$ operators}
\label{AppendixEth}
 In this appendix, the operators $\bm\partial_{+}$ and $\bm\partial_{-}$ are
written in terms of the $\eth$ and $\bar{\eth}$ operators of Newman and Penrose.
 To fix the notation
and conventions, let $\eth_{P}$ and $\bar{\eth}_{P}$ denote the $\eth$ and $\bar{\eth}$
operators  \cite{PenRin84} as defined in
\cite{Ste91}.
In the language of the
 NP-formalism \cite{NewPen62,PenRin84,Ste91}, given a null frame
 represented by $\{\bml,\bmn,\bmm,\bar{\bmm}\}$ their
corresponding covariant directional derivatives are denoted by
$\{D,\Delta,\delta,\bar{\delta}\}$.  The
operators $\eth_{P}$ and $\bar{\eth}_{P}$ acting on a quantity $\eta$
with spin weight $s$ can be written in terms of the $\delta$ and
$\bar{\delta}$ derivatives as ---see \cite{Ste91},
\begin{equation}
\label{EthandEthBarStewart}
\eth_{P} \eta = \delta \eta + s(\bar{\alpha}-\beta)\eta, \qquad
\bar{\eth}_{P}\eta=\bar{\delta}\eta-s(\alpha-\bar{\beta})\eta,
\end{equation}
where $\alpha$ and $\beta$ denote the spin coefficients as defined
in the NP formalism.
 The action of the directional derivatives $ \delta$ and
 $\bar{\delta}$ on the vectors $\bmm$ and $\bar{\bmm}$, projected into
 the tangent space $T(\mathcal{Q}) \subset T(\mathcal{M})$ spanned by
 $\bmm$ and $\bar{\bmm}$, is encoded in
\begin{equation}
\label{SpinCoefsOnSphereStewart}
\delta m^a=-(\bar{\alpha}-\beta)m^{a}, \qquad
\delta{\bar{m}^a}=(\bar{\alpha}-\beta)\bar{m}^a \qquad \text{on}\qquad
\mathcal{Q}.
\end{equation}
The directional derivatives $\slashed{\nabla}_{+}$
and $\slashed{\nabla}_{-}$ as defined on Appendix
\ref{ConnectionTwoSphere} are related to $\delta$ and $\bar{\delta}$
via
\[
\delta=\frac{1}{\sqrt{2}}\slashed{\nabla}_{+}, \qquad
\bar{\delta}=\frac{1}{\sqrt{2}}\slashed{\nabla}_{-}.
\]
It follows from the discussion of Appendix \ref{ConnectionTwoSphere}
and equation \eqref{SpinCoefsOnSphereStewart} that
\begin{equation}\label{SpinCoefsOnSphereGasVal}
\bar{\alpha}-\beta=-\frac{1}{\sqrt{2}}\overline{\varpi}, \qquad
\text{on} \qquad \mathcal{Q}
\end{equation}
Using equations \eqref{SpinCoefsOnSphereStewart} and
\eqref{SpinCoefsOnSphereGasVal} one obtains
\begin{equation}\label{DirectionalDerivativesToEthandEthbar}
\slashed{\nabla}_{+}\eta=\sqrt{2}\eth_{P}\eta +
s\overline{\varpi}\eta, \qquad
\slashed{\nabla}_{-}\eta=\sqrt{2}\bar{\eth}_{P}\eta - s\varpi\eta,
\end{equation}
To align the discussion with the conventions of \cite{FriKan00, Val07b, 
Val03a} is convenient to define $\eth$ and $\bar{\eth}$ by rescaling
$\eth_{P}$ and $\bar{\eth}_{P}$ as
\begin{equation}\label{EthAndEthBarForUs}
\eth\equiv-\frac{1}{\sqrt{2}}\eth_{P}, \qquad
\bar{\eth}\equiv-\frac{1}{\sqrt{2}}\bar{\eth}_{P}.
\end{equation}
The corresponding eigenfunctions $Y_{s; \ell m}$ of the operator
$\eth\bar{\eth}$, defining the spin-weighted spherical harmonics, will
be assumed to be rescaled in accordance with equation
\eqref{EthAndEthBarForUs}.  Exploiting that $\{Y_{s; \ell m} \}$, with
$0\leq |s| \leq \ell$ and $-\ell \leq m \leq \ell$, form a complete
basis for functions of spin-weight s over $\mathbb{S}^2$, given a
scalar field $\xi:\mathcal{Q} \rightarrow \mathbb{R}$,  with
spin-weight $s$, one can expand $\xi$ as
\begin{equation}
\label{GeneralExpansionSphericalHarmonics}
\xi=\sum_{\ell=s}^{\infty}\sum_{m=-\ell}^{\ell}C_{s\ell m} \; Y_{s; \ell m}.
\end{equation}  
In addition, one has that
\begin{subequations}
\begin{align}
& \eth(Y_{s;\ell m})=\sqrt{(\ell-s)(\ell+ s +1)} \; Y_{s+1;\ell,m},
 \label{RiseSpin}\\
& \bar{\eth}(Y_{s;\ell,m})=-\sqrt{(\ell + s)(\ell-s +1)} 
\; Y_{s-1;\ell m}. \label{LowerSpin}
\end{align} 
\end{subequations}
Notice that equation \eqref{GeneralExpansionSphericalHarmonics} as
well as equations \eqref{RiseSpin}-\eqref{LowerSpin} do not depend on
the specific choice of coordinates on $\mathcal{Q}$.

% 
%\bibliographystyle{reporthack}
% Ludovica
%\bibliographystyle{/Users/Juan/Documents/tex/reporthack}

% Path in QM 
%\bibliography{ThesisGRbib}
% Path in Ludovica
%\bibliography{/Users/Juan/Documents/tex/Newgrbib}

\end{document}